\documentclass[a4paper,fleqn,10pt]{article}
\pdfoutput=1
\usepackage{amsmath}

\usepackage{amssymb}
\usepackage{array}
\usepackage{calc}
\usepackage{longtable}
\usepackage{multirow}
\usepackage{graphicx}
\usepackage{xspace}
\usepackage{todonotes}
\presetkeys{todonotes}{inline}{}

\usepackage{mciteplus}
\usepackage[a4paper,pdfborder={0 0 0}]{hyperref}
\usepackage[format=hang,labelfont=bf,hypcap=true]{caption}
\usepackage{subcaption}
\usepackage{sectsty}
\allsectionsfont{\sffamily}
\subsubsectionfont{\mdseries\itshape\large}
\makeatletter
\DeclareRobustCommand*{\bfseries}{%
  \not@math@alphabet\bfseries\mathbf
  \fontseries\bfdefault\selectfont
  \boldmath
}
\makeatother
\let\spreprint\empty
\newcommand{\preprint}[1]{\def\spreprint{\protect#1}}
\let\sinstitute\empty
\newcommand{\institute}[1]{\def\sinstitute{\protect#1}}
\makeatletter
\renewcommand{\maketitle}{\begingroup
  \null\thispagestyle{empty}%
    \ifx\spreprint\empty
      \vskip 5ex
    \else
      \flushright\large\spreprint\vskip 2ex
    \fi
    \vskip 5ex
    \flushleft
      {\sffamily\bfseries\huge\@title}\vskip 6ex
      \@author\vskip 2ex
      \ifx\sinstitute\empty
      \else
        {\small\sinstitute}
      \fi
    \vskip 3ex
  \endgroup
}
\makeatother
\renewenvironment{abstract}{\begin{center}
  {\large\sffamily\bfseries Abstract: }
  \begin{minipage}[t]{0.75\textwidth}
}{\end{minipage}\end{center}\vskip 10ex}

\numberwithin{equation}{section}

\newcommand{\Herwig}{H\protect\scalebox{0.8}{ERWIG}\xspace}

\newcommand{\Ariadne}{A\protect\scalebox{0.8}{RIADNE}\xspace}

\newcommand{\Pythia}{P\protect\scalebox{0.8}{YTHIA}\xspace}

\newcommand{\Rivet}{R\protect\scalebox{0.8}{IVET}\xspace}

\newcommand{\Sherpa}{S\protect\scalebox{0.8}{HERPA}\xspace}
\newcommand{\Dire}{D\protect\scalebox{0.8}{IRE}\xspace}

\long\def\symbolfootnote[#1]#2{\begingroup%
\def\thefootnote{\fnsymbol{footnote}}\footnote[#1]{#2}\endgroup}
\newcommand{\abs}[1]{\left| #1\right|}
\newcommand{\rbr}[1]{\left( #1\right)}

\newcommand{\sbr}[1]{\left[ #1\right]}
\newcommand{\im}{\imath}
\newcommand{\jm}{\jmath}
\newcommand{\done}{{\rm d}}
\newcommand{\order}{\mathcal{O}}
\newcommand{\mc}[1]{\mathcal{#1}}

\newcommand{\wt}[1]{\widetilde{#1}}

\hypersetup{
  pdfauthor={Stefan Hoeche,Stefan Prestel}
  pdftitle={The midpoint between dipole and parton showers}
}
\author{Stefan H{\"o}che$^1$, Stefan Prestel$^1$}
\title{The midpoint between dipole and parton showers}
\institute{
  $^1$ SLAC National Accelerator Laboratory, 
  Menlo Park, CA 94025, USA}
\preprint{SLAC-PUB-16304\\MCNET-15-13}
\begin{document}
\maketitle
\begin{abstract}
  We present a new parton-shower algorithm. Borrowing from the basic ideas of dipole
  cascades, the evolution variable is judiciously chosen as the transverse momentum 
  in the soft limit. This leads to a very simple analytic structure of the evolution.
  A weighting algorithm is implemented, that allows to consistently treat potentially
  negative values of the splitting functions and the parton distributions. We provide
  two independent, publicly available implementations for the two event generators
  \Pythia and \Sherpa.
\end{abstract}
\section{Introduction}
Parton showers and fragmentation models have been used for more than three decades
to predict the dynamics of multi-particle final states in collider
experiments~\cite{Webber:1986mc,*Buckley:2011ms}. More recently, the traditional 
approaches implemented in \Herwig~\cite{Marchesini:1987cf,*Gieseke:2003rz},
\Pythia~\cite{Sjostrand:1985xi,*Sjostrand:2004ef} and \Sherpa~\cite{Kuhn:2000dk,*Krauss:2005re}
were supplemented by methods based on dipole and antenna factorization~\cite{Nagy:2006kb,*Nagy:2014mqa,
  Schumann:2007mg,*Dinsdale:2007mf,Winter:2007ye,Giele:2007di,*Giele:2011cb,*Ritzmann:2012ca,Platzer:2009jq}.
A characteristic feature of these new shower programs is the description of QCD coherence
in the color dipole picture~\cite{Gustafson:1986db}, which has first been implemented 
in the \Ariadne Monte Carlo~\cite{Gustafson:1987rq,*Lonnblad:1992tz,*Kharraziha:1997dn}.
In this article we present a dipole-like parton shower similar to existing ones, 
but we focus on the simplest implementation and enforce sum rules and DGLAP collinear 
anomalous dimensions. We choose ordering variables based on transverse momenta 
in the soft approximation, while existing dipole-like shower models employ collinear transverse momenta.
As such, the model is a hybrid of dipole and parton shower.
These choices will eventually allow to compare with analytic approaches, such as 
CSS~\cite{Collins:1981uk,*Collins:1984kg,*Collins:1985ue,*Sterman:1986aj,*Collins:1988ig,*Collins:1989gx}
and SCET~\cite{Bauer:2000yr,*Bauer:2001ct,*Bauer:2001yt,*Bauer:2002nz}.

In the past decade, the matching of parton showers to NLO calculations~\cite{
  Frixione:2002ik,*Nason:2004rx,*Frixione:2007vw,*Frixione:2010ra,*Torrielli:2010aw,
  *Alioli:2010xd,*Hoeche:2010pf,*Hoeche:2012ft,*Platzer:2011bc,*Alwall:2014hca,*Jadach:2015mza,
  Hoeche:2011fd} 
and the merging of LO~\cite{Catani:2001cc,*Lonnblad:2001iq,*Mangano:2001xp,*Alwall:2007fs,
  *Lavesson:2007uu,*Hamilton:2009ne,*Hamilton:2010wh,*Hoeche:2010kg,*Lonnblad:2012ng,*Platzer:2012bs}
and NLO matched results for different jet multiplicity~\cite{
  Lavesson:2008ah,*Gehrmann:2012yg,Hoeche:2012yf,*Lonnblad:2012ix,*Frederix:2012ps,*Alioli:2012fc}
was in the focus of interest of the majority of Monte-Carlo developers~\cite{Nason:2012pr}. 
Comparably few efforts were made to provide publicly available implementations of parton-showers%
~\cite{Nagy:2006kb,*Nagy:2014mqa,Schumann:2007mg,Giele:2007di,*Giele:2011cb,Platzer:2009jq}
or to improve their formal accuracy~\cite{Jadach:2011cr,*Jadach:2013dfd}, and even fewer of the 
new parton showers have made their way into complete event generators used by experiments.
When comparing results of matched and merged calculations, it is therefore often unclear 
whether a particular difference stems from mismodeling in the parton shower, 
from differences in the matching or merging algorithm, or simply from technical problems.
Similarly, when comparing the results of different event generators at the hadron level
it is often unclear whether differences should be ascribed to the hadronization model,
to the simulation of multiple scattering / rescattering effects, or to the parton shower.
We intend to remedy this situation to some extent, by providing two implementations of 
one and the same algorithm, to be used with the two different event generation frameworks
\Pythia~\cite{Sjostrand:2014zea} and \Sherpa\cite{Gleisberg:2003xi,*Gleisberg:2008ta}.
We subject our codes to rigorous scrutiny by comparing their
predictions at the sub-permille level.

This paper is organized as follows: Section~\ref{sec:ps_formalism} reviews the basic
parton-shower formalism. Section~\ref{sec:dire} explains the construction principles
of our new parton shower, which we call \Dire (acronym for DIpole REsummation).
Section~\ref{sec:validation} contains the validation of the numerical implementation,
and Sec.~\ref{sec:results} presents a comparison of the predictions from \Dire with
experimental measurements. Section~\ref{sec:conclusions} contains some concluding remarks.

\section{Parton-shower formalism}
\label{sec:ps_formalism}
The evolution of parton densities and fragmentation functions in the collinear limit 
is governed by the DGLAP equations~\cite{Gribov:1972ri,*Dokshitzer:1977sg,*Altarelli:1977zs}:
\begin{equation}\label{eq:pdf_evolution}
  \frac{\done f_{a}(x,t)}{\done\ln t}=
  \sum_{b=q,g}\int_x^1\frac{\done z}{z}\,\frac{\alpha_s}{2\pi}\sbr{P_{ba}(z)}_+\,f_{b}(x/z,t)\;,
\end{equation}
where $P_{ab}$ are the regularized evolution kernels. Assume that we define
$P_{ab}$ in terms of unregularized kernels, $\hat{P}_{ab}$, restricted to all
but an $\varepsilon$-environment around the soft-collinear pole, plus an endpoint contribution.
\begin{equation}\label{eq:sf_regularization}
  \begin{split}
    P_{ba}(z,\varepsilon)=&\;\hat{P}_{ba}(z)\,\Theta(1-z-\varepsilon)
    -\delta_{ab}\,\frac{\Theta(z-1+\varepsilon)}{\varepsilon}
    \sum_{c=q,g}\int_0^{1-\varepsilon}\done\zeta\,\zeta\,\hat{P}_{ac}(\zeta)\\
    =&\;\hat{P}_{ba}(z)\,\Theta(1-z-\varepsilon)
    +\delta_{ab}\,\frac{\Theta(z-1+\varepsilon)}{\varepsilon}\,
    \Big(2C_a\,\ln\varepsilon+\gamma_a+\order(\varepsilon)\Big)\;.
  \end{split}
\end{equation}
For finite $\varepsilon$, the endpoint subtraction can be interpreted as the approximate virtual plus 
unresolved real corrections, which are included in the parton shower by enforcing unitarity.
The precise value of $\varepsilon$ is defined in terms of an infrared cutoff on the evolution variable,
using four-momentum conservation. When ignoring momentum conservation, this cutoff can be taken to zero,
which allows us to identify $\sbr{P_{ba}(z)}_+$ as the $\varepsilon\to 0$ limit of $P_{ba}(z,\varepsilon)$.
For $0<\varepsilon\ll 1$, Eq.~\eqref{eq:pdf_evolution} changes to
\begin{equation}\label{eq:pdf_evolution_constrained}
  \frac{1}{f_{a}(x,t)}\,\frac{\done f_{a}(x,t)}{\done\ln t}=
  -\sum_{c=q,g}\int_0^{1-\varepsilon}\done\zeta\,\zeta\,\frac{\alpha_s}{2\pi}\hat{P}_{ac}(\zeta)\,
  +\sum_{b=q,g}\int_x^{1-\varepsilon}\frac{\done z}{z}\,
  \frac{\alpha_s}{2\pi}\,\hat{P}_{ba}(z)\,\frac{f_{b}(x/z,t)}{f_{a}(x,t)}\;.
\end{equation}
Using the Sudakov form factor
\begin{equation}
  \Delta_a(t_0,t)=\exp\bigg\{-\int_{t_0}^{t}\frac{\done \bar{t}}{\bar{t}}
  \sum_{c=q,g} \int_0^{1-\varepsilon}\done\zeta\,\zeta\,\frac{\alpha_s}{2\pi}\hat{P}_{ac}(\zeta)\bigg\}
  \approx\exp\bigg\{-\int_{t_0}^{t}\frac{\done \bar{t}}{\bar{t}}\frac{\alpha_s}{2\pi}\Big[
    2C_a\ln\frac{1}{\varepsilon{\scriptstyle (t_0,t)}}-\gamma_a\Big]\bigg\}
\end{equation}
one can define the generating functional for splittings of parton $a$ as
\begin{equation}\label{eq:def_updf}
  \mc{F}_a(x,t,\mu^2)=f_a(x,t)\Delta_a(t,\mu^2)=f_a(x,\mu^2)\,\Pi_a(x,t,\mu^2)\;,
\end{equation}
where
\begin{equation}\label{eq:def_pi}
  \Pi_a(x,t_0,t)=\exp\bigg\{-\int_{t_0}^{t}\frac{\done \bar{t}}{\bar{t}}
  \sum_{b=q,g}\int_x^{1-\varepsilon}\frac{\done z}{z}\,
  \frac{\alpha_s}{2\pi}\,\hat{P}_{ba}(z)\,\frac{f_{b}(x/z,\bar{t})}{f_{a}(x,\bar{t})}\bigg\}\;.
\end{equation}
In this context, $\Pi_a(x,t,\mu^2)$ is the probability that the parton does not 
undergo a branching process between the two scales $\mu^2$ and $t$~\cite{Ellis:1991qj}.
Equation~\eqref{eq:pdf_evolution_constrained} can now be written in the simple form
\begin{equation}\label{eq:pdf_evolution_constrained_2}
  \frac{\done\ln\mc{F}_a(x,t,\mu^2)}{\done\ln t}
  =\sum_{b=q,g}\int_x^{1-\varepsilon}\frac{\done z}{z}\,
  \frac{\alpha_s}{2\pi}\,\hat{P}_{ba}(z)\,\frac{f_{b}(x/z,t)}{f_{a}(x,t)}\;.
\end{equation}
The generalization to an $n$-parton state can involve multiple PDFs and fragmentation functions:
\begin{equation}\label{eq:pdf_evolution_constrained_3}
  \begin{split}
  \frac{\done\ln\mc{F}_{\vec{a}}(\hat{\Phi}_n,t,\mu^2)}{\done\ln t}
  =&\sum_{i\in{\rm IS}}\sum_{b=q,g}\int_{x_i}^{1-\varepsilon}\frac{\done z}{z}\,
  \frac{\alpha_s}{2\pi}\,\hat{P}_{ba_i}(z)\,\frac{f_{b}(x_i/z,t)}{f_{a_i}(x_i,t)}\\
  &\qquad+\sum_{j\in{\rm FS}}\sum_{b=q,g}\int_{z_j}^{1-\varepsilon}\frac{\done z}{z}\,
  \frac{\alpha_s}{2\pi}\,\hat{P}_{a_jb}(z)\,\frac{D_{b}(z_j/z,t)}{D_{a_j}(z_j,t)}\;.
  \end{split}
\end{equation}
In this context, we have extended the argument of the generating functional
to $\hat{\Phi}_n$, which denotes the $n$-parton phase-space configuration, including all
light-cone momentum fractions, $x_i$ and $z_j$, for  initial-state (IS) and final-state (FS)
partons. $\mc{F}$ also depends on all parton flavors, denoted by $\vec{a}$.
If we do not fix the momenta of final-state hadrons, the fragmentation functions 
can be integrated over $z_j$, leading to the simplified formula
\begin{equation}\label{eq:pdf_evolution_constrained_4}
  \frac{\done\ln\mc{F}_{\vec{a}}(\Phi_n,t,\mu^2)}{\done\ln t}
  =\sum_{i\in{\rm IS}}\sum_{b=q,g}\int_{x_i}^{1-\varepsilon}\frac{\done z}{z}\,
  \frac{\alpha_s}{2\pi}\,\hat{P}_{ba_i}(z)\,\frac{f_{b}(x_i/z,t)}{f_{a_i}(x_i,t)}
  +\sum_{j\in{\rm FS}}\sum_{b=q,g}\int_\varepsilon^{1-\varepsilon}\done z\,
  \frac{\alpha_s}{2\pi}\,\hat{P}_{a_jb}(z)\;.
\end{equation}
The change from $\hat{\Phi}_n$ to $\Phi_n$ signals that $\mc{F}$ has become independent of $z_j$.
An observable-dependent generating functional for the parton shower can now be defined recursively as
\begin{equation}\label{eq:shower_functional}
  \begin{split}
    \mc{F}_{\vec{a}}(\Phi_n,t,t';O)=&\;\mc{F}_{\vec{a}}(\Phi_n,t,t')\,O(\Phi_n)
    +\int_t^{t'}\frac{\done\bar{t}}{\bar{t}}\,
    \frac{\done\ln\mc{F}_{\vec{a}}(\Phi_n,\bar{t},t')}{\done\ln\bar{t}}\,
    \mc{F}_{\vec{a}'}(\Phi'_{n+1},t,\bar{t};O)\;.
  \end{split}
\end{equation}
The first term includes all virtual corrections and unresolved real emissions, resummed into 
a no-branching probability. The second term describes a single branching, followed by
further parton evolution. Both terms can be generated simultaneously by implementing the 
veto algorithm~\cite{Sjostrand:2006za} for Eq.~\eqref{eq:def_pi}. 
We have introduced an observable, $O$, that measures the kinematics
of the final state. In general, this observable will act differently on the no-emission term 
and on the emission term. In the trivial case that $O=1$, Eq.~\eqref{eq:shower_functional} 
returns the unitarity constraint, $\mc{F}_{\vec{a}}(\Phi_n,t,t';1)=1$.

Generating a branching in the parton shower involves selecting a new color topology for the
$n+1$-particle state. For non-trivial color configurations, $\mc{F}$ will therefore depend on 
the color assignment in the large-$N_c$ limit. While it is in principle necessary to keep track 
of this dependence, we omit any notation relating to color in order to simplify our final formulae. 
The selection of color topologies proceeds as in existing dipole-like parton showers, which is
described in great detail in~\cite{Schumann:2007mg}. It is straightforward to extend our notation
in this regard.

The choice of evolution variable is crucial. At leading color the soft radiation
pattern emerges from the coherent gluon radiation off ``color dipoles'' that are spanned by the
two partons at opposite ends of a color string~\cite{Andersson:1983ia,*Bassetto:1984ik}.
This mandates the choice of an evolution variable which treats these two partons democratically.
In other words, the evolution variable should be identical no matter whether one or the other
of two color-connected partons is considered the radiator. This will be the guiding principle
for its selection in Sec.~\ref{sec:dire}.

The splitting functions, $P(z)$, are formally defined in the collinear limit, and they do not
reflect the soft radiation pattern outside the collinear region. Traditionally, this problem 
is dealt with by imposing angular ordering constraints on the final-state phase space~\cite{
  Marchesini:1987cf,*Gieseke:2003rz}. Alternatively one can use the approach of Catani and 
Seymour~\cite{Catani:1996vz}, and introduce a $t$-dependence in the splitting functions that
restores the correct soft anomalous dimension at one-loop order~\cite{Schumann:2007mg,*Dinsdale:2007mf,Platzer:2009jq}.
We will use this procedure in the next section. It is important that the modified splitting 
functions satisfy the sum rules, which are enforced by Eq.~\eqref{eq:sf_regularization}
and by the corresponding flavor sum rule~\cite{Ellis:1991qj}. The new splitting functions 
may also be negative in the non-singular phase-space region. This requires a modification 
of the Sudakov veto algorithm~\cite{Hoeche:2009xc,Platzer:2011dq,*Lonnblad:2012hz,Hoeche:2011fd}, 
and it entails an analytic event weight to ensure that both emission- and no-emission probabilities 
are accounted for. We find that, in our parton-shower approach, the variance of this weight is small. 
In fact, for both final-final and initial-initial dipoles momentum conservation guarantees that 
no negative weights can arise from the splitting functions. Negative weights may however appear 
in initial-initial configurations due to negative values of the PDFs.

\section{Construction of the \texorpdfstring{\Dire}{Dire} shower}
\label{sec:dire}
\begin{figure}[t]
  \begin{center}
    \includegraphics[width=0.8\textwidth]{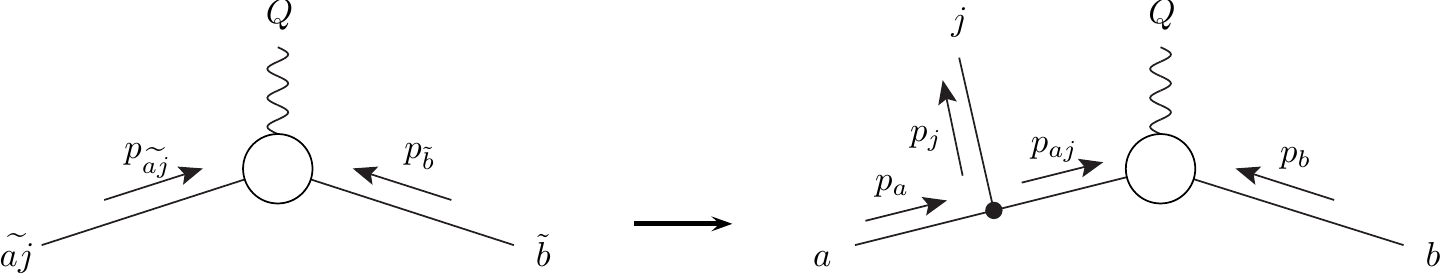}
    \caption{Kinematics in the initial-state parton splitting process $a\to \{aj\}j$.
      \label{fig:split_kinematics}}
  \end{center}
\end{figure}
A basic branching process is sketched in Fig.~\ref{fig:split_kinematics}. In this case we consider
initial-state evolution. We employ the kinematics from Ref.~\cite{Catani:1996vz,Catani:2002hc},
which we summarize in Appendix~\ref{sec:kinematics}. For initial-state splitters with initial-state
spectator, all particles typically have zero on-shell mass, which greatly simplifies the calculation. 
Their momenta can be parametrized in terms of the light-cone momenta $p_a$ and $p_b$, 
using the standard Sudakov decomposition~\cite{Sudakov:1954sw}
\begin{align}\label{eq:sud_decomposition}
  p_{aj}^\mu=\;&z\,p_a^\mu+\frac{\vec{k}_\perp^2-2\,p_ap_j}{z}\frac{p_b^\mu}{2p_ap_b}+k_\perp^\mu\;,
  &p_j^\mu=\;&(1-z)\,p_a^\mu+\frac{\vec{k}_\perp^2}{1-z}\frac{p_b^\mu}{2p_ap_b}-k_\perp^\mu\;.
\end{align}
In the Catani-Seymour approach~\cite{Catani:1996vz}, the correct soft anomalous dimension
is obtained after replacing the soft enhanced term in the splitting functions by a partial fraction
of the soft eikonal for the color dipole defined by the splitting parton and its spectator.
Schematically this can be done as follows:
\begin{equation}\label{eq:cs_replacement}
  \frac{1}{1-z}\to\frac{1}{1-z+v}
  \qquad\text{where}\qquad
  1-z=\frac{p_jp_b}{p_ap_b}\;,\qquad v=\frac{p_jp_a}{p_ap_b}\;.
\end{equation}
If we define the evolution variable of our parton shower to be a scaled transverse momentum, 
$t=(z-v)\,\vec{k}_\perp^2$, then the soft-enhanced term in Eq.~\eqref{eq:cs_replacement} 
is conveniently expressed as
\begin{equation}\label{eq:def_ii_kt}
  \frac{1-z}{(1-z)^2+\kappa^2},
  \qquad\text{where}\qquad
  \kappa^2=\frac{t}{Q^2}=\frac{p_ap_j\;p_jp_b}{(p_ap_b)^2}\;,
  \qquad
  Q^2=2\,p_ap_b-2\,(p_a+p_b)p_j\;.
\end{equation}
Note that the evolution variable has the desired symmetry property, i.e.\ it is symmetric in 
emitter and spectator momentum. More precisely, our evolution variable is the exact inverse
of the soft eikonal. As such, it is different from the hardness parameter, $k_\perp^2$.
Consequently, the parton shower will fill the entire final-state phase space, even for 
factorization scales much smaller than the hadronic center of mass energy.

We define the evolution kernels using the replacement of the soft enhanced term in
Eq.~\eqref{eq:cs_replacement}. Additionally, we require the collinear anomalous
dimension to be unchanged. Imposing flavor and momentum sum rules, these two requirements
determine the complete set of leading-order spin-averaged splitting functions as:
\begin{equation}\label{eq:ap_kernels}
  \begin{split}
    P_{qq}(z,\kappa^2)=\;&2\,C_F\bigg[\left(\frac{1-z}{(1-z)^2+\kappa^2}\right)_+-\frac{1+z}{2}\bigg]+\frac{3}{2}\,C_F\,\delta(1-z)\\
    P_{gg}(z,\kappa^2)=\;&2\,C_A\bigg[\left(\frac{1-z}{(1-z)^2+\kappa^2}\right)_++\frac{z}{z^2+\kappa^2}-2+z(1-z)\bigg]
    +\delta(1-z)\left(\frac{11}{6}C_A-\frac{2}{3}n_f T_R\right)\\
    P_{qg}(z,\kappa^2)=\;&2\,C_F\bigg[\frac{z}{z^2+\kappa^2}-\frac{2-z}{2}\bigg]\\
    P_{gq}(z,\kappa^2)=\;&T_R\bigg[z^2+(1-z)^2\bigg]
  \end{split}
\end{equation}
The corresponding anomalous dimensions are listed in Appendix~\ref{sec:anom_dim}.
Using the phase-space factorization derived in~\cite{Catani:1996vz}, we obtain the following
differential branching probability:
\begin{equation}\label{eq:pi_ii}
  \frac{\done\ln\mc{F}^{\rm(II)}_a(x,t,\mu^2)}{\done\ln t}
  =\sum_{b=q,g}\int_{z_-}^{z_+}\frac{\done z}{z-v}\,
  \frac{\alpha_s(t)}{2\pi}\,\hat{P}_{ba}(z)\,\frac{f_{b}(x/(z-v),t)}{f_{a}(x,t)}\;,
\end{equation}
where $2\,z_\pm=1+x\pm\sqrt{(1-x)^2-4\kappa^2}$.\footnote{
  The actual value of the integration limits on $z$ does not have to be computed explicitly.
  In practice, one generates Monte-Carlo events in the maximum range $x<z<1$, and vetoes events 
  violating momentum conservation (cf.~\cite{Sjostrand:2006za}).} Note that this implies $x/(z-v)<1$,
i.e.\ the light-cone momentum fraction entering the PDFs is well defined. This variable
has changed compared to Eq.~\eqref{eq:pdf_evolution} to account for four-momentum 
conservation, while Eq.~\eqref{eq:pdf_evolution} remains valid in the collinear limit, $v\to 0$.
In addition, the strong coupling is evaluated at the evolution variable, hence the Landau pole 
is avoided by the infrared cutoff of the parton shower, $t_0$ (which is of order 1~GeV).
Finally, the splitting kinematics are constructed as described in App.~\ref{sec:css_kin_ii}.

The technical implementation of Eqs.~\eqref{eq:ap_kernels} in terms of dipole terms proceeds 
as described in~\cite{Schumann:2007mg}. That is, we divide the splitting function according 
to the number of spectator partons in the large-$N_c$ limit, and sum over color-adjacent 
splitter-spectator pairs. The corresponding evolution equations are straightforward extensions 
of Eq.~\eqref{eq:pi_ii}, and therefore we do not list them here.
The same reasoning applies to all dipole types discussed in the following.

Initial-state splitters with final-state spectator are treated along the same lines. 
The construction of final-state momenta is described in App.~\ref{sec:css_kin_if}.
The kinematics are now defined in terms of the following variables
($p_k$ is the final-state spectator momentum)
\begin{equation}\label{eq:def_if_kin}
  z=1-\frac{p_jp_k}{p_ap_j+p_ap_k}\;,\qquad
  u=\frac{p_jp_a}{p_ap_j+p_ap_k}\;.
\end{equation}
We define the evolution variable, $t$, and its dimensionless variant, $\kappa^2$, as
\begin{equation}\label{eq:def_if_kt}
  t=Q^2\,u\,(1-z)\;,\qquad
  \kappa^2=\frac{t}{Q^2}=\frac{p_ap_j\;p_jp_k}{(p_ap_j+p_ap_k)^2}\;,
  \qquad\text{where}\qquad
  Q^2=2\,p_a(p_j+p_k)-2\,p_jp_k\;.
\end{equation}
The evolution variable is symmetric in emitter and spectator momentum, up to a rescaling by $1/z^2$.
The differential branching probability is:
\begin{equation}\label{eq:pi_if}
  \frac{\done\ln\mc{F}^{\rm(IF)}_a(x,t,\mu^2)}{\done\ln t}
  =\sum_{b=q,g}\int_x^{1-t/Q^2}\frac{\done z}{z}\,
  \frac{\alpha_s(t)}{2\pi}\,\hat{P}_{ba}(z)\,\frac{f_{b}(x/z,t)}{f_{a}(x,t)}\;.
\end{equation}
Final-state splitters with initial-state spectator must have the same evolution variable 
due to symmetry arguments. Therefore the asymmetric rescaling by $1/z^2$ in the IF case must also 
be applied in the FI case. The only choice to be made concerns the splitting variable, 
which is taken to be $1-u$. The differential branching probability is:
\begin{equation}\label{eq:pi_fi}
  \frac{\done\ln\mc{F}^{\rm(FI)}_a(x,t,\mu^2)}{\done\ln t}
  =\sum_{b=q,g}\int_{z_-}^{z_+}\done z\,
  \frac{\alpha_s(t)}{2\pi}\,\hat{P}_{ab}(z)\,\frac{f_{a}(x/\bar{z},t)}{f_{a}(x,t)}\,
  \Theta\big(Q^2\,(1-z)(1-x)-t\big)\;,
\end{equation}
where $\bar{z}=1-t/Q^2/(1-z)$.
The integration limits are given by $z_-=1-t/Q^2/(1-x)$ and $z_+=1-t/Q^2$.
The construction of final-state momenta is described in App.~\ref{sec:css_kin_fi}.\\
For final-state splittings, $ij\to i,j$, with final-state spectator, $k$, we use the variables
\begin{equation}\label{eq:def_csvars_ff}
  y=\frac{2\,p_ip_j}{Q^2}\;,\qquad
  \tilde{z}=\frac{p_ip_k}{p_ip_k+p_jp_k}\;,
  \qquad\text{where}\qquad
  Q^2=2\,p_ip_k+2\,(p_i+p_k)p_j\;.
\end{equation}
The symmetric evolution variable, its scaled variant $\kappa^2$ and the splitting variable
are defined as
\begin{equation}\label{eq:def_ff_kt}
  t=Q^2\,y(1-y)\,(1-\tilde{z})\;,\qquad
  \kappa^2=\frac{t}{Q^2}=\frac{2\,p_ip_j\;2\,p_jp_k}{Q^4}\;,\qquad
  z=1-(1-\tilde{z})(1-y)\;.
\end{equation}
The differential branching probability is:
\begin{equation}\label{eq:pi_ff}
  \frac{\done\ln\mc{F}^{\rm(FF)}_a(t,\mu^2)}{\done\ln t}
  =\sum_{b=q,g}\int_{z_-}^{z_+}\done z\,\frac{\alpha_s(t)}{2\pi}\,\hat{P}_{ab}(z)\;,
\end{equation}
where $2\,z_\pm=1\pm\sqrt{1-4\,t_0/Q^2}$.
The splitting kinematics are described in App.~\ref{sec:css_kin_ff}.

Note that the scaled transverse momentum defined in Eq.~\eqref{eq:def_ff_kt} is substantially different 
from the ones defined in~\cite{Schumann:2007mg,*Dinsdale:2007mf,Platzer:2009jq}, which can be written as
\begin{equation}\label{eq:collinear_kt}
  \tilde{\kappa}^2=\frac{\tilde{t}}{Q^2}=\frac{2\,p_ip_j}{Q^2}\frac{p_ip_k\,p_kp_j}{((p_i+p_j)p_k)^2}\;.
\end{equation}
This variable is symmetric in $i$ and $j$, but not in $i$ and $k$, which would be required
in order to interpret it as the inverse soft eikonal for gluon radiation off the dipole spanned 
by $i$ and $k$. Kinematically, Eq.~\eqref{eq:collinear_kt} represents the transverse momentum 
of partons $i$ / $j$ with respect to the anti-collinear direction defined by $k$.
This is what we call a ``collinear'' transverse momentum. In contrast, Eq.~\eqref{eq:def_ff_kt}
can be interpreted as the transverse momentum of the two daughter dipoles $(ij)$ / $(kj)$ in the 
center-of-mass frame of the decaying dipole~\cite{Gustafson:1987rq,*Lonnblad:1992tz}. In this case, 
$i$ defines the collinear, and $k$ defines the anti-collinear direction, making the symmetry explicit. 
We refer to such a definition as a ``soft'' transverse momentum.

The change in the definition of transverse momenta compared to existing $p_T$-ordered dipole-like
parton showers~\cite{Schumann:2007mg,Platzer:2009jq} also involves changing the splitting variable, 
in order to reduce the related Jacobians to unity while maintaining Eqs.~\eqref{eq:ap_kernels}, 
simultaneously for all dipole types. In contrast, the kinematics mapping is identical to the 
previously published methods~\cite{Schumann:2007mg,Carli:2009cg}.

If massive quarks are involved in the branching process, we would like to map the evolution
variable to the soft enhanced term of the full matrix element, just like in the massless limit.
The singularity structure in the soft limit is given in~\cite{Catani:2002hc}.
For the most involved case of two massive radiators, $i$ and $k$, it leads to an eikonal of the form
\begin{equation}\label{eq:massive_eikonal}
  \frac{p_ip_k}{p_ip_j\;p_jp_k}-\frac{m_i^2}{2\,(p_ip_j)^2}-\frac{m_k^2}{2\,(p_kp_j)^2}\;.
\end{equation}
The inverse of this function is difficult to interpret. Its scaling property in the 
soft limit, however, is completely determined by the first term in Eq.~\eqref{eq:massive_eikonal},
whose inverse can therefore be used to define an ordering variable for the evolution of massive partons
\begin{equation}\label{eq:massive_kt}
  t=\frac{2\,p_ip_j\;2\,p_jp_k}{2\,p_ip_k}=k_\perp^{(0)\,2}+
  \left(m_i^2\zeta_i^2+m_k^2\zeta_k^2\right)
  \frac{\gamma(s_{ik},m_i^2,m_k^2)}{s_{ik}-m_i^2-m_k^2}\;,
\end{equation}
with $\gamma$ defined in Appendix~\ref{sec:kinematics}, and with $s_{ik}=(p_i+p_k)^2$.
Here we have defined the massless equivalent of the evolution variable, $k_\perp^{(0)\,2}$,
and the light-cone momentum fractions, $\zeta_i$ and $\zeta_j$ in a Sudakov decomposition
of the gluon momentum, $p_j$, along the directions of $p_i$ and $p_k$:
\begin{equation}
  p_j=p_i\left(\zeta_i-\frac{m_k^2\zeta_k}{\gamma(s_{ik},m_i^2,m_k^2)}\right)+
  p_k\left(\zeta_k-\frac{m_i^2\zeta_i}{\gamma(s_{ik},m_i^2,m_k^2)}\right)+k_\perp^{(0)}\;.
\end{equation}
Equation~\eqref{eq:massive_kt} is valid in the soft limit. For practical purposes
the denominator $p_ip_k$ in the evolution variable should be the hard scale of the radiating dipole,
which is given by $(2\,p_ap_b)^2/Q^2$, $(2\,p_a(p_i+p_k))^2/Q^2$ and $Q^2$ for II, IF/FI and FF dipoles, respectively.

The splitting functions for massive partons can be taken from Eq.~\eqref{eq:ap_kernels} 
and be modified according to~\cite{Catani:2002hc}. We use the following unregularized 
massive kernels for final-state splitter with final- or initial-state spectator:
\begin{equation}\label{eq:massive_kernels_f}
  \begin{split}
    \hat{P}^{\rm(F)}_{QQ}(z,\kappa^2)=\;&C_F\bigg[\,2\,\frac{1-z}{(1-z)^2+\kappa^2}
      -\frac{v_{\wt{\im\jm},\tilde{k}}}{v_{ij,k}}\bigg(1+z+\frac{m_Q^2}{p_Qp_g}\bigg)\bigg]\\
    \hat{P}^{\rm(F)}_{gg}(z,\kappa^2)=\;&2\,C_A\bigg[\,\frac{1-z}{(1-z)^2+\kappa^2}+\frac{z}{z^2+\kappa^2}
      -\frac{2-z(1-z)}{v_{ij,k}}\,\bigg]\\
    \hat{P}^{\rm(F)}_{gQ}(z,\kappa^2)=\;&T_R\,\frac{1}{v_{ij,k}}\,\bigg[1-2\,z(1-z)
      +\frac{m_Q^2}{p_{\vphantom{\bar{Q}}Q}p_{\bar{Q}}+m_Q^2}\bigg]\;.
  \end{split}
\end{equation}
The relative velocity between two momenta, $p$ and $q$, is defined as
\begin{equation}
  v_{p,q}=\frac{\beta((p+q)^2,p^2,q^2)}{(p+q)^2-p^2-q^2}=\sqrt{1-\frac{p^2q^2}{(pq)^2}}\;,
\end{equation}
and $v_{\wt{\im\jm},\tilde{k}}$ and $v_{ij,k}$ stand for the relative velocities between
the emitter parton and the spectator before and after the branching, respectively.
The branching probabilities are modified as $\hat{P}_{ab}(z,\kappa^2)\to J(z,\kappa^2)\,\hat{P}_{ab}(z,\kappa^2)$,
where $J(z,t)$ is a spectator-dependent Jacobian factor~\cite{Catani:2002hc,Schumann:2007mg}.
It is unity, except for the case of final-state splitter with final-state spectator, where
\begin{equation}
  J^{\rm(FF)}(y)=\frac{Q^2}{\sqrt{\lambda(Q^2+m_i^2+m_j^2+m_k^2,m_{ij}^2,m_k^2)}}
  \bigg(1+\frac{m_i^2+m_j^2-m_{ij}^2}{Q^2\,y}\bigg)^{-1}\;,
\end{equation}
using $Q^2=2\,p_ip_k+2\,(p_i+p_k)p_j$, cf.\ Eq.~\eqref{eq:def_csvars_ff}.
The phase-space boundaries are given by the roots of the Gram determinant
\begin{equation}\label{eq:gram_3}
  4\Delta_3=2p_ip_j\,2p_jp_k\,2p_ip_k
  -(2p_ip_j)^2m_k^2-(2p_jp_k)^2m_i^2-(2p_ip_k)^2m_j^2
  +4m_i^2m_j^2m_k^2\;.
\end{equation}
While the massless case leads to simple constraints on $z$, the general massive case
generates a rather involved functional form of the $z$-boundary as a function of $t$. 
Algorithmically, it is preferable to use the veto algorithm~\cite{Sjostrand:2006za}
to implement this constraint, or to use Eqs.~\eqref{eq:def_ff_kt} and~\eqref{eq:def_if_kt} 
and evaluate the constraint in collinear variables, where~\cite{Catani:2002hc}
\begin{equation}\label{eq:massive_z_bounds}
  \tilde{z}_\pm=\,\frac{p_ip_j+m_i^2}{(p_i+p_j)^2}\,
      \big(\,1\pm v_{ij,j}v_{ij,k}\,\big)\;.
\end{equation}
In final-state splittings with initial-state spectator the PDF is evaluated at
$x/\bar{z}/(1+(m_{ij}^2-m_i^2-m_j^2)/Q^2)$. Correspondingly, the theta function in Eq.~\eqref{eq:pi_fi}
changes to $\Theta\big(Q^2\,(1-z)(1-x\,Q^2/(Q^2+m_{ij}^2-m_i^2-m_j^2))-t\big)$.\\
For initial-state splitter with final-state spectator the mass-dependent splitting functions are
\begin{equation}\label{eq:massive_kernels_i}
  \begin{split}
    \hat{P}^{\rm(I)}_{qg}(z,\kappa^2)=\;&C_F\,\bigg[\,
      2\,\frac{z}{z^2+\kappa^2}-(2-z)
      -\frac{2\,m_k^2}{Q^2}\frac{u}{1-u}\bigg]\\
    \hat{P}^{\rm(I)}_{gg}(z,\kappa^2)=\;&2\,C_A\bigg[\,
      \frac{1-z}{(1-z)^2+\kappa^2}+\frac{z}{z^2+\kappa^2}-2+z(1-z)
      -\frac{m_k^2}{Q^2}\frac{u}{1-u}\,\bigg]\;.
  \end{split}
\end{equation}

\section{Validation}
\label{sec:validation}
In this section we validate the numerical implementation of the \Dire parton shower.
The two event generation frameworks \Pythia~\cite{Sjostrand:1985xi,*Sjostrand:2004ef} and
\Sherpa~\cite{Kuhn:2000dk,*Krauss:2005re} are used to construct two entirely independent
Monte Carlo programs. Aside from a thorough cross-check of the implementation, this allows,
for the first time, to simulate Deep Inelastic Scattering in \Pythia~8.
We employ the CT10nlo PDF set~\cite{Lai:2010vv}, and the corresponding
value of the strong coupling. Following standard practice to improve the logarithmic accuracy
of the parton shower, the soft enhanced term of the splitting functions is rescaled
by $1+\alpha_s(t)/(2\pi) K$, where $K=(67/18-\pi^2/6)\,C_A-10/9\,T_R\,n_f$~\cite{Catani:1990rr}.

\begin{figure}[t]
  \centering
  \begin{minipage}{7.5cm}
    \includegraphics[width=\textwidth]{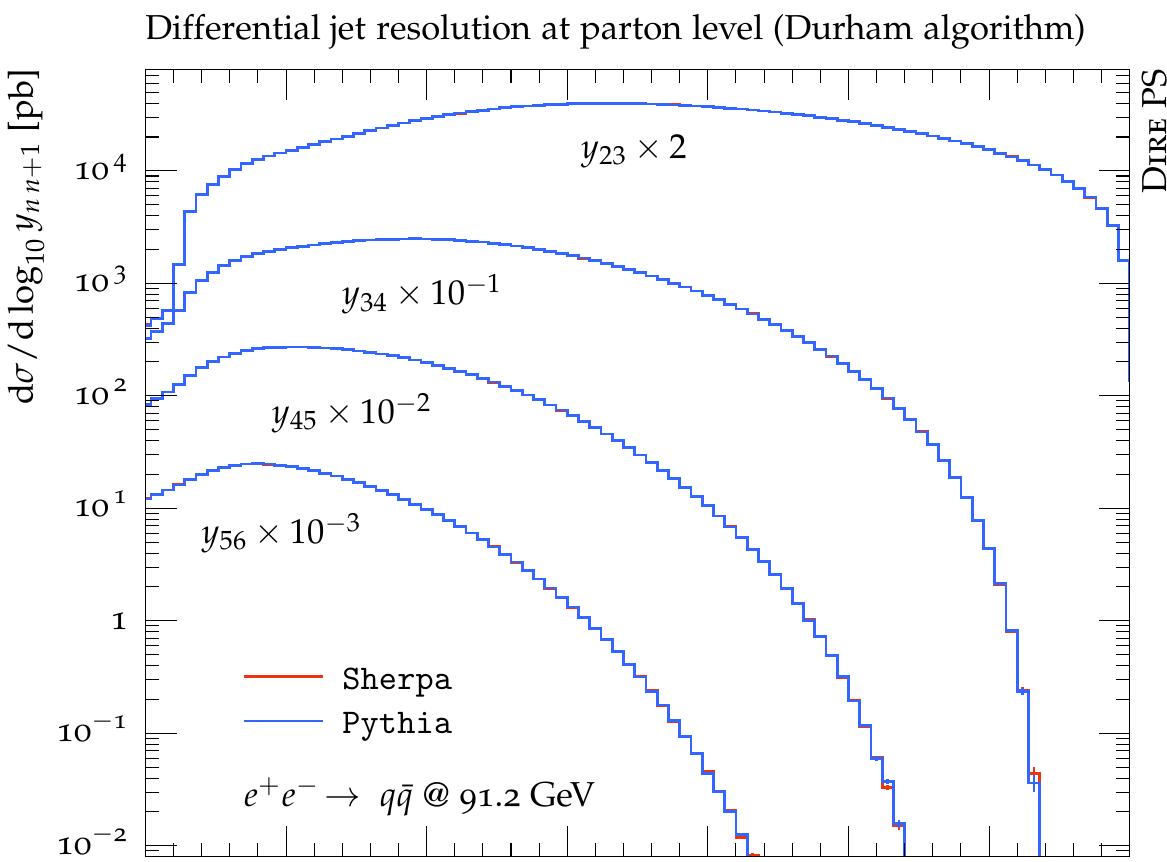}\\[-1mm]
    \includegraphics[width=\textwidth]{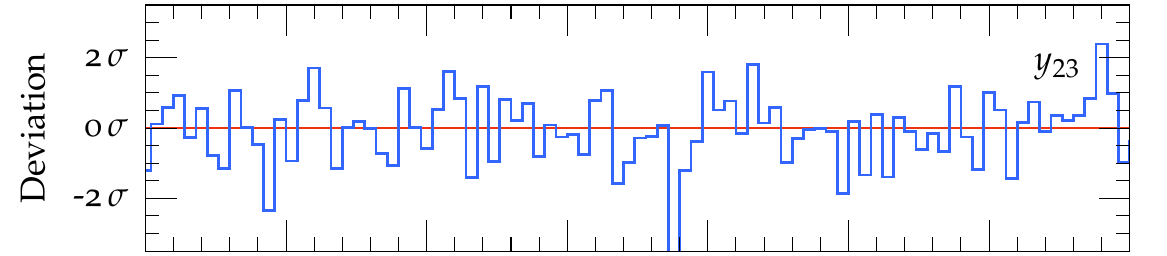}\\[-1mm]
    \includegraphics[width=\textwidth]{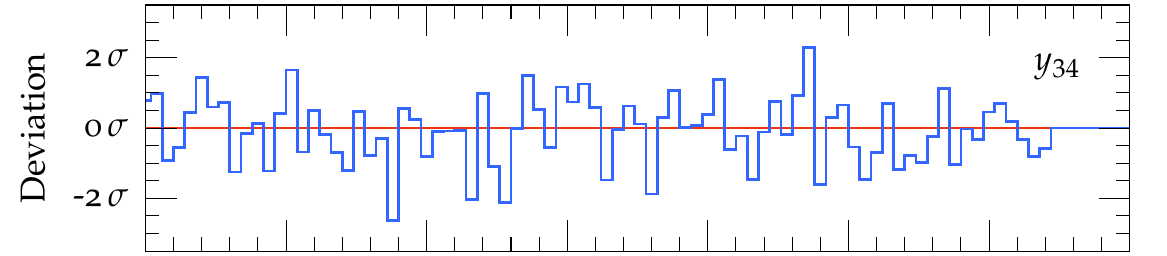}\\[-1mm]
    \includegraphics[width=\textwidth]{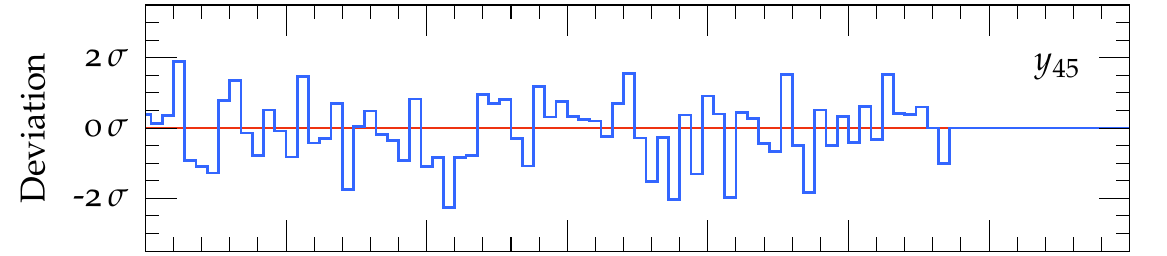}\\[-1mm]
    \includegraphics[width=\textwidth]{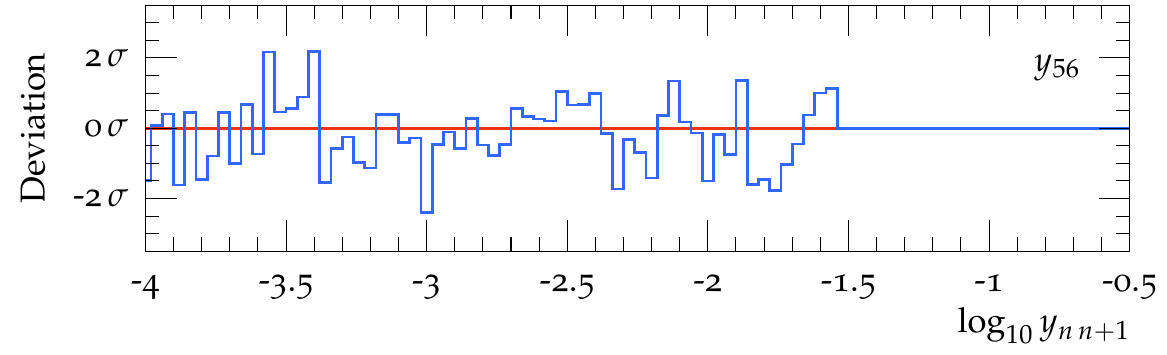}
  \end{minipage}\hskip 5mm
  \begin{minipage}{7.5cm}
    \includegraphics[width=\textwidth]{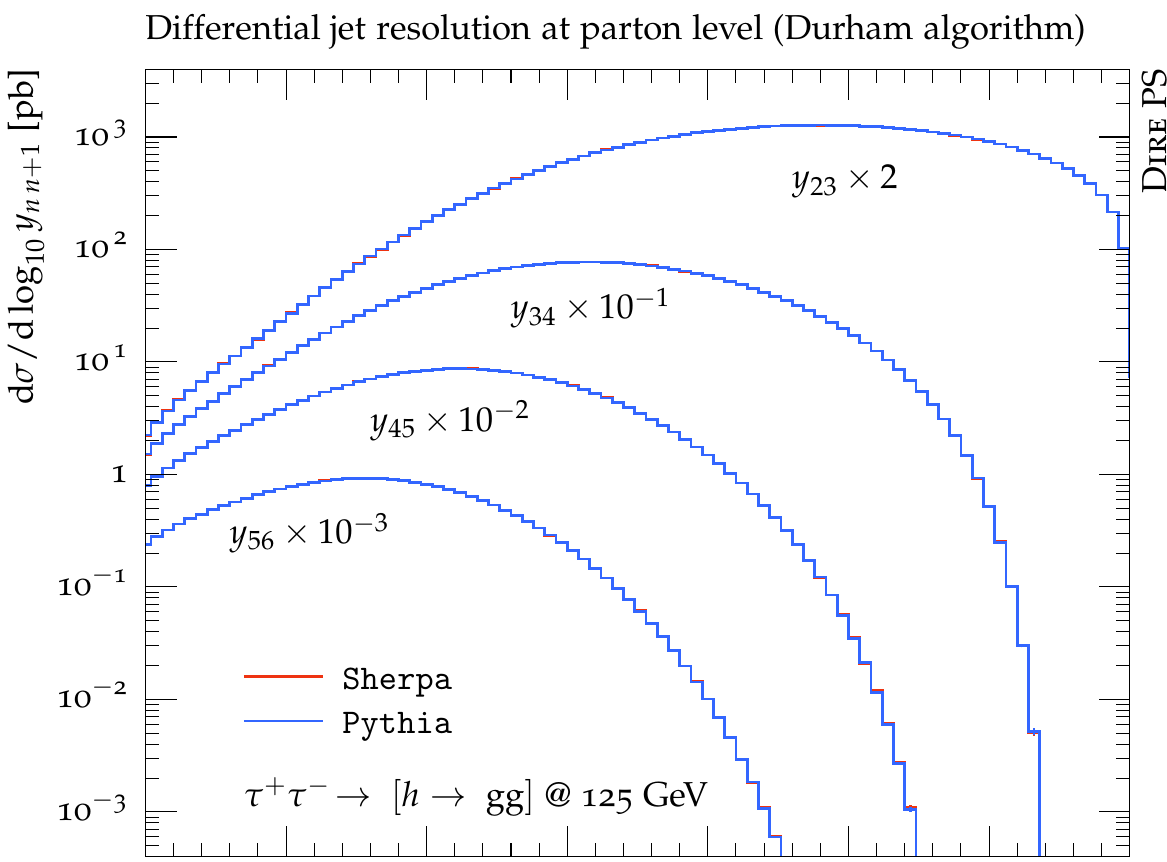}\\[-1mm]
    \includegraphics[width=\textwidth]{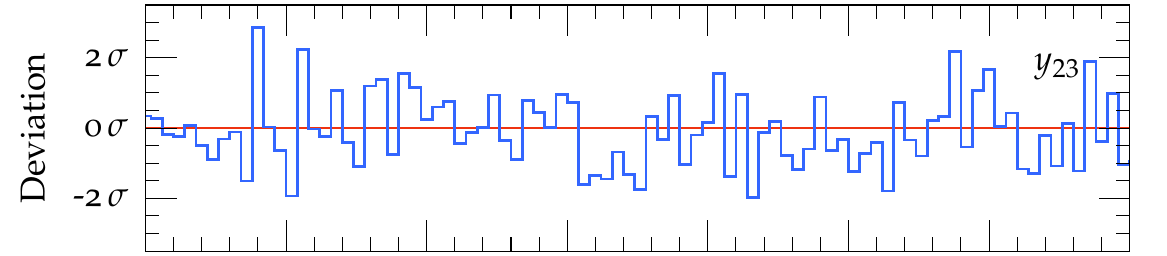}\\[-1mm]
    \includegraphics[width=\textwidth]{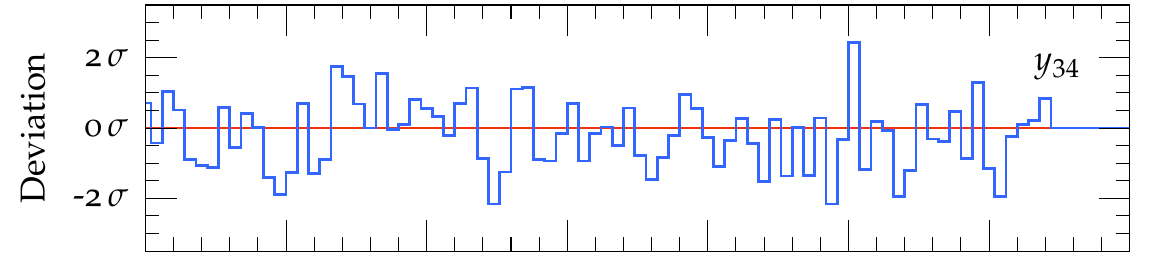}\\[-1mm]
    \includegraphics[width=\textwidth]{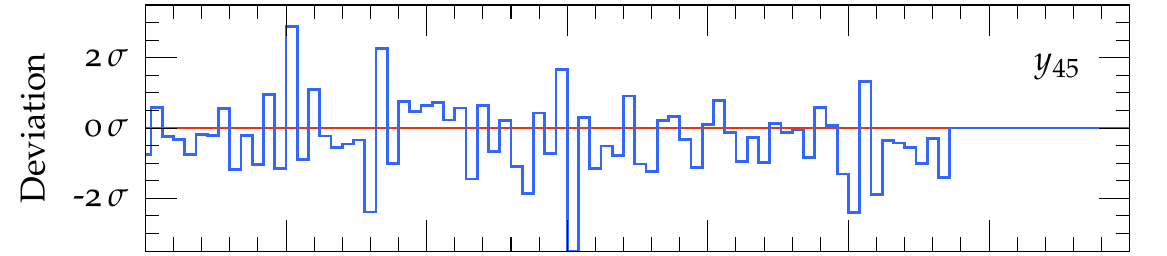}\\[-1mm]
    \includegraphics[width=\textwidth]{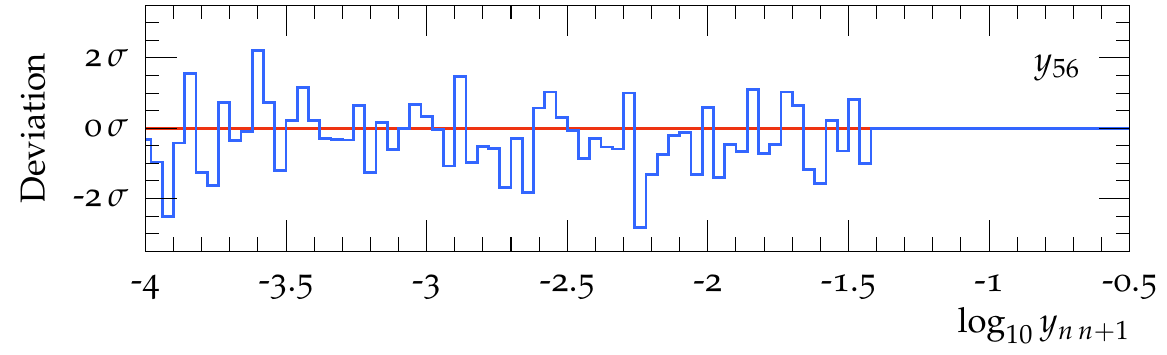}
  \end{minipage}
  \caption{Validation in $e^+e^-\to\text{hadrons}$ and $\tau^+\tau^-\to[h\to\text{hadrons}]$
    \label{fig:validation_ee}}
\end{figure}

\begin{figure}[p]
  \centering
  \begin{minipage}{7.5cm}
    \includegraphics[width=\textwidth]{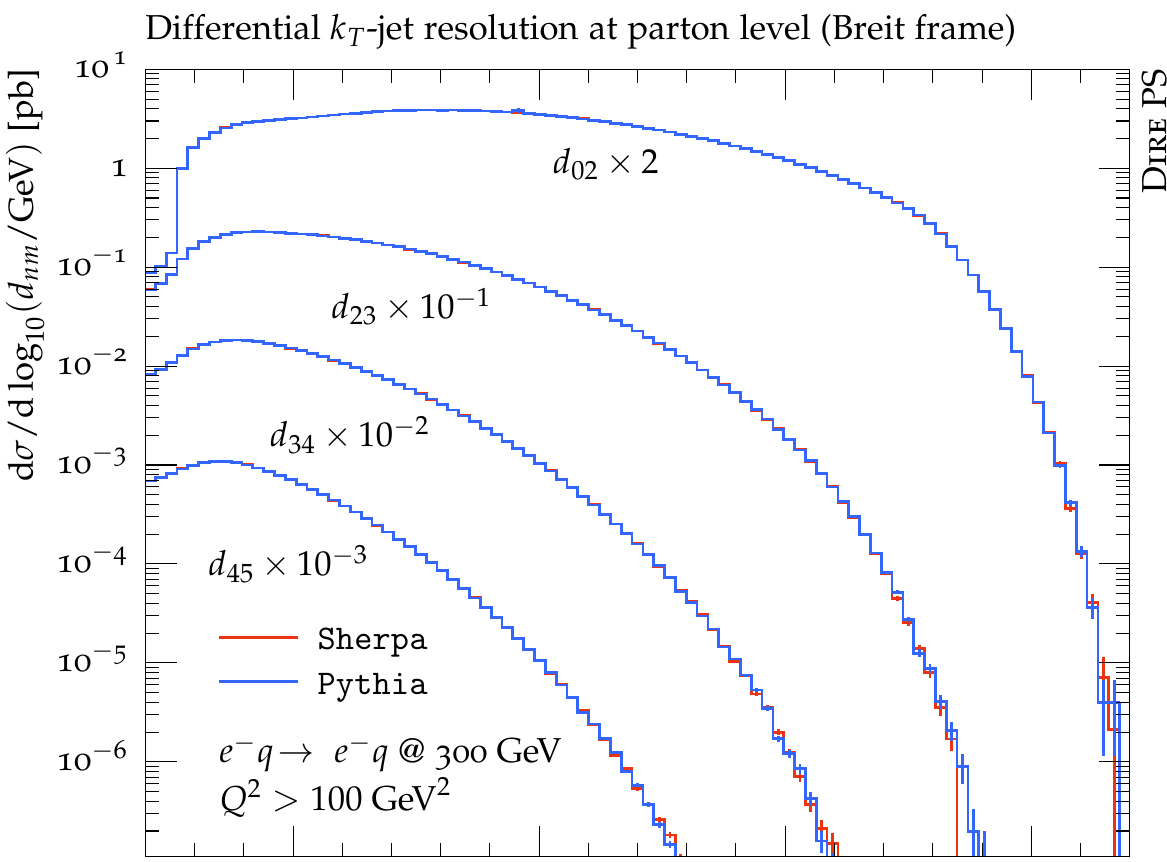}\\[-1mm]
    \includegraphics[width=\textwidth]{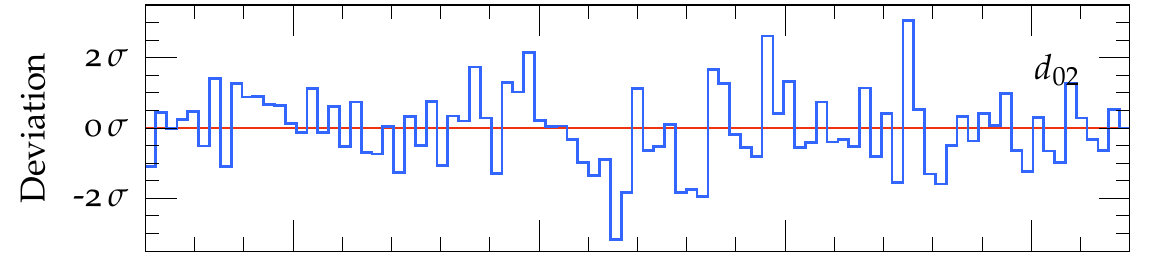}\\[-1mm]
    \includegraphics[width=\textwidth]{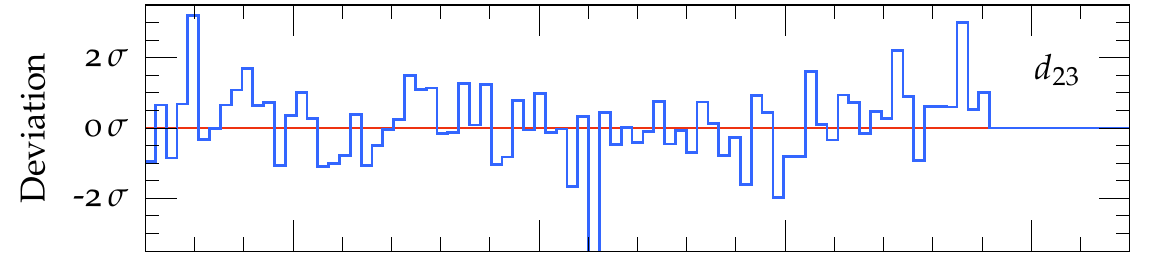}\\[-1mm]
    \includegraphics[width=\textwidth]{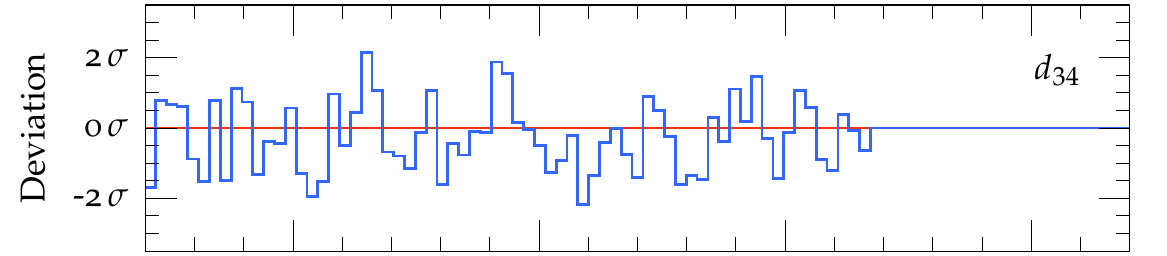}\\[-1mm]
    \includegraphics[width=\textwidth]{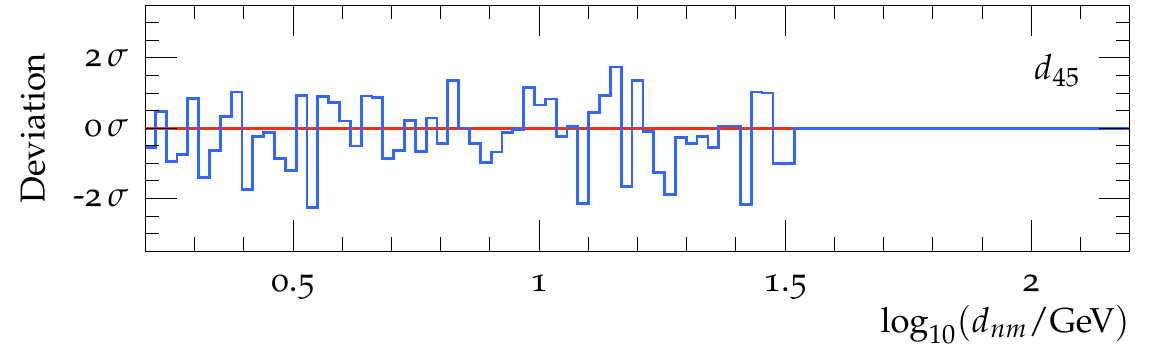}
  \end{minipage}\hskip 5mm
  \begin{minipage}{7.5cm}
    \includegraphics[width=\textwidth]{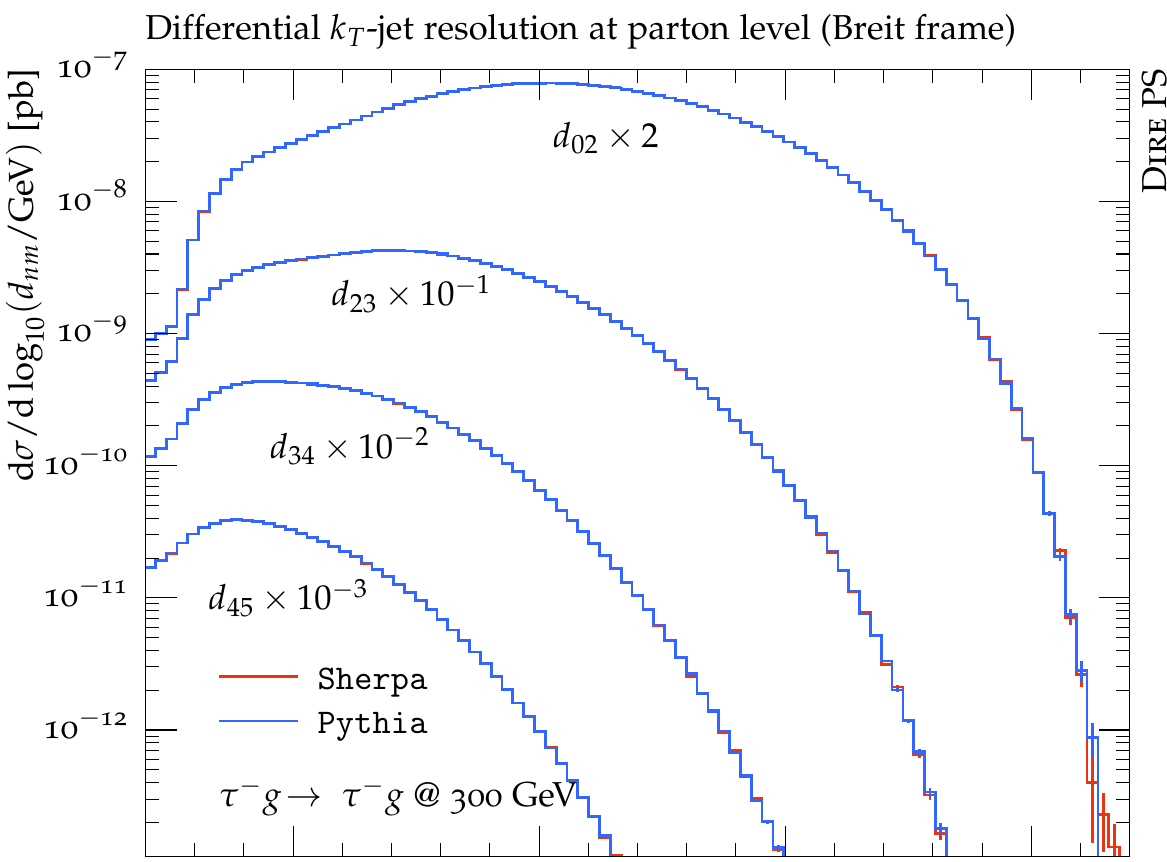}\\[-1mm]
    \includegraphics[width=\textwidth]{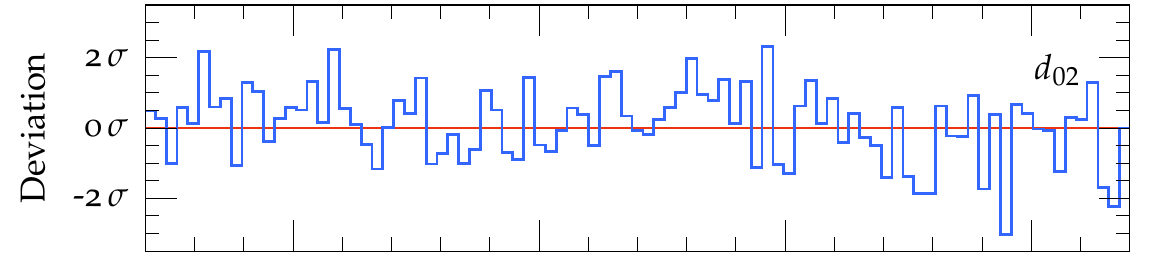}\\[-1mm]
    \includegraphics[width=\textwidth]{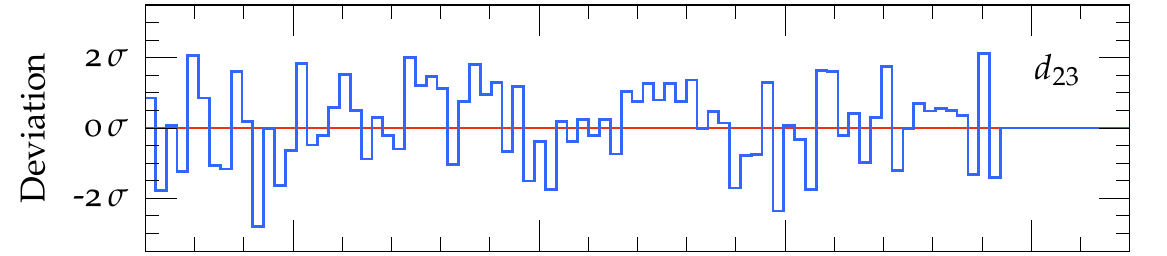}\\[-1mm]
    \includegraphics[width=\textwidth]{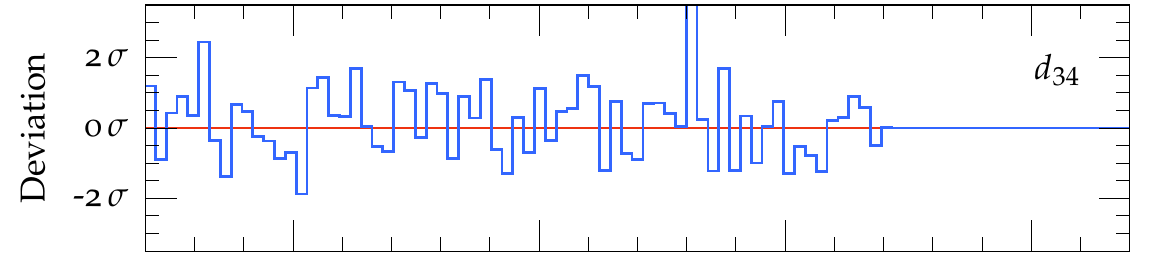}\\[-1mm]
    \includegraphics[width=\textwidth]{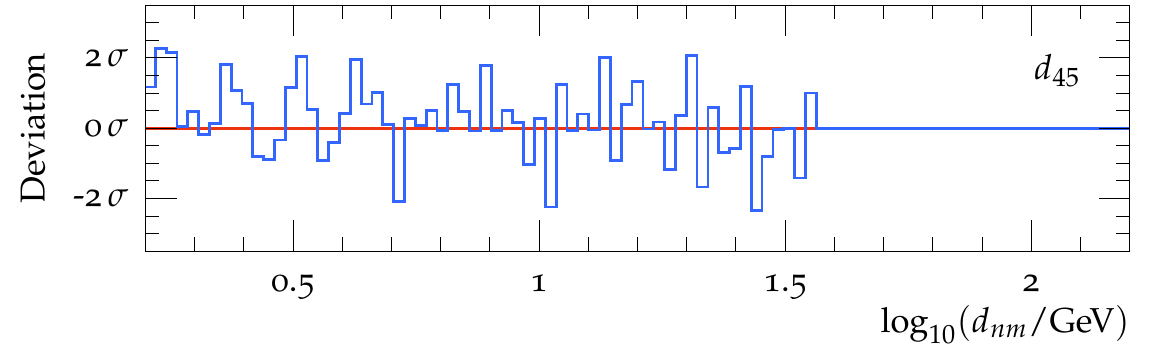}
  \end{minipage}\hskip 5mm
  \caption{Validation in $e^-q\to e^-q$ and $\tau^-g\to\tau^-g$
    \label{fig:validation_ep}}
\end{figure}

\begin{figure}[p]
  \centering
  \begin{minipage}{7.5cm}
    \includegraphics[width=\textwidth]{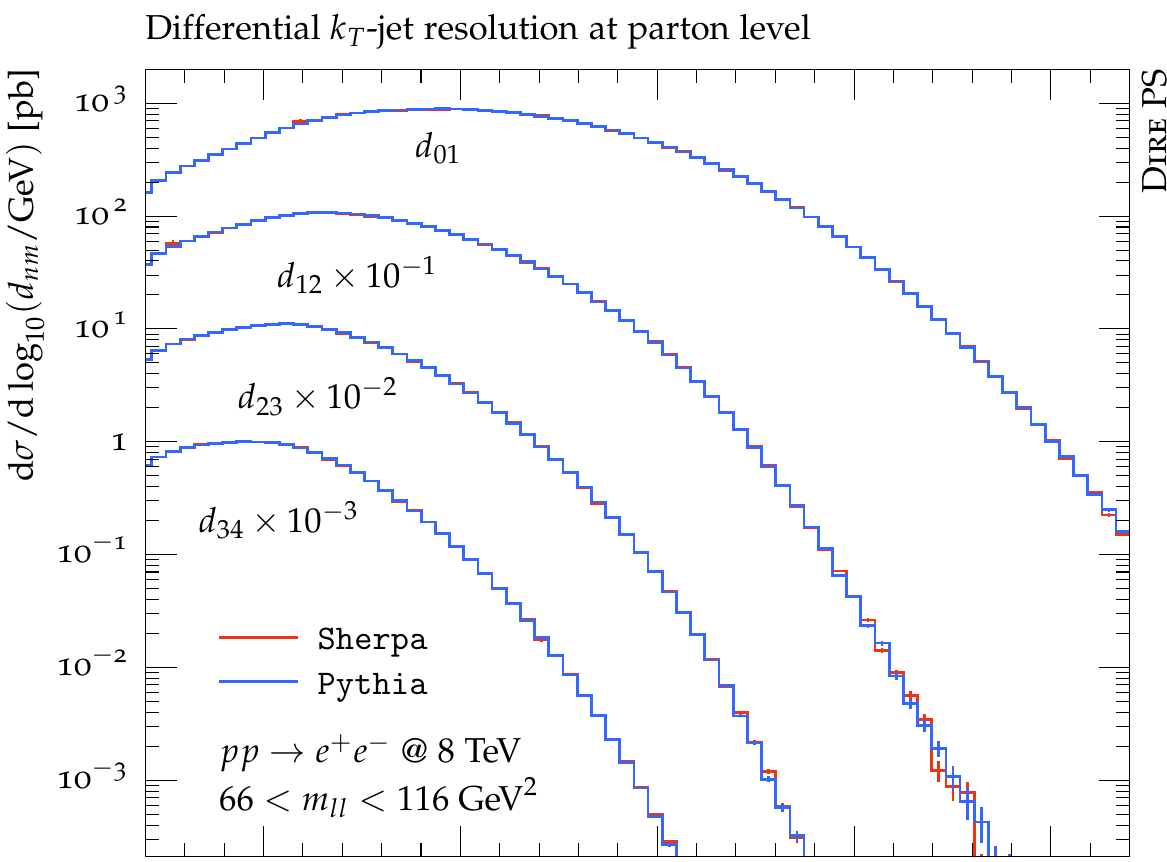}\\[-1mm]
    \includegraphics[width=\textwidth]{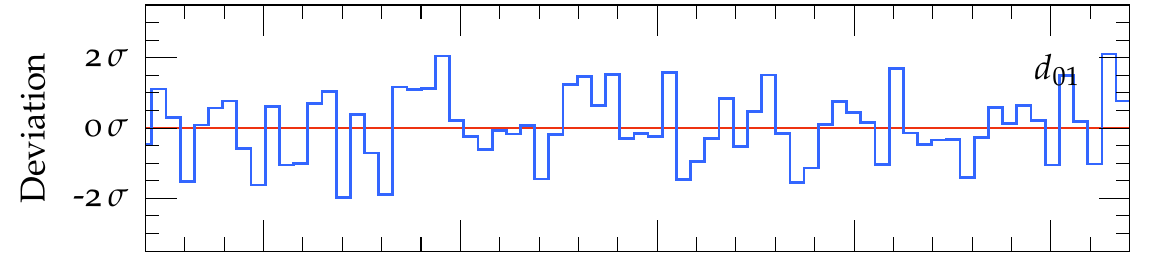}\\[-1mm]
    \includegraphics[width=\textwidth]{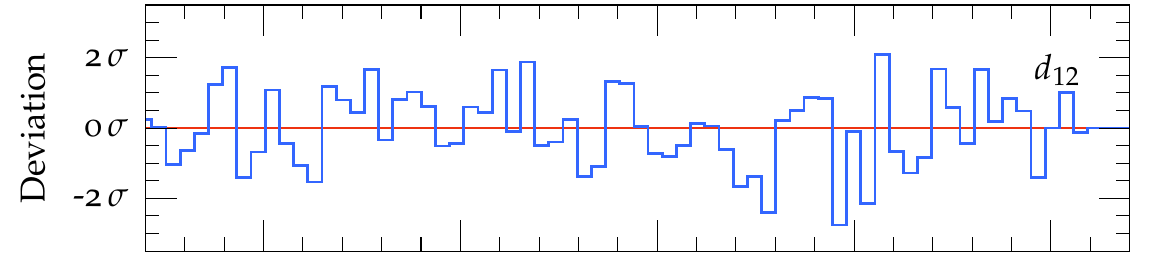}\\[-1mm]
    \includegraphics[width=\textwidth]{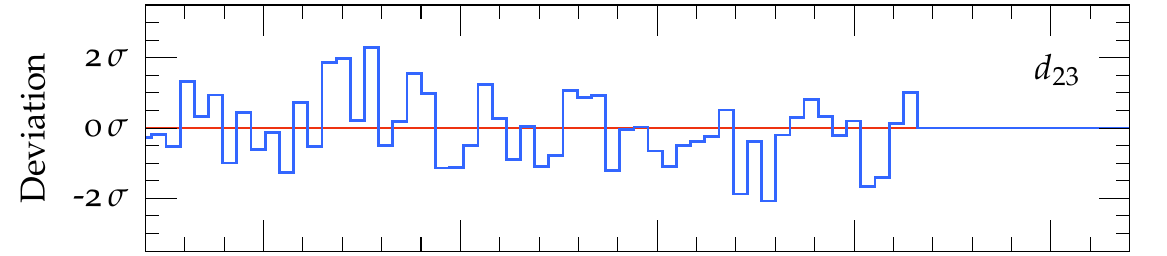}\\[-1mm]
    \includegraphics[width=\textwidth]{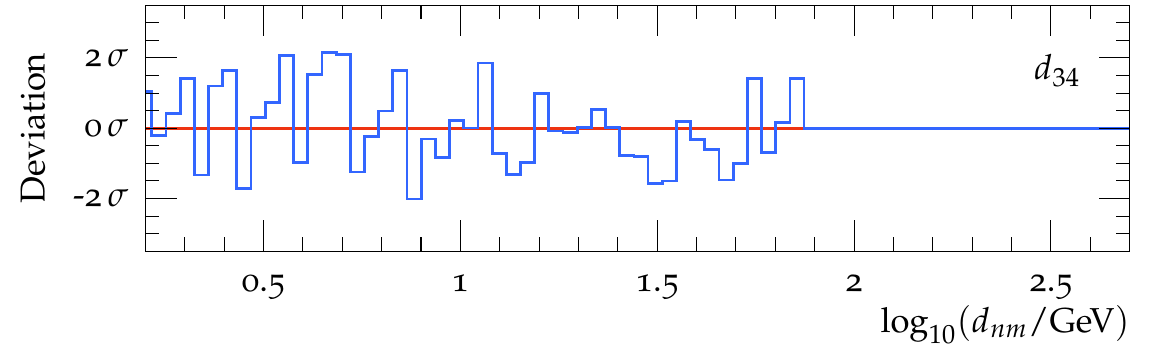}
  \end{minipage}\hskip 5mm
  \begin{minipage}{7.5cm}
    \includegraphics[width=\textwidth]{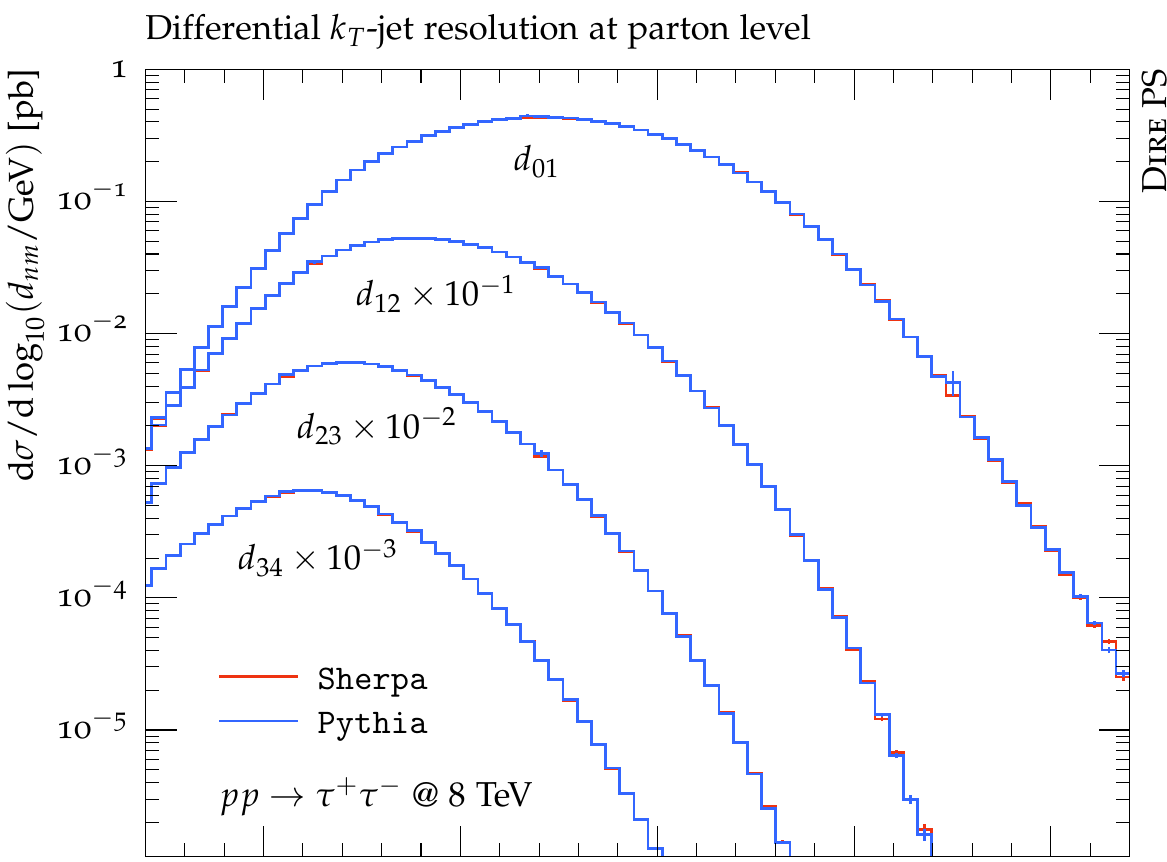}\\[-1mm]
    \includegraphics[width=\textwidth]{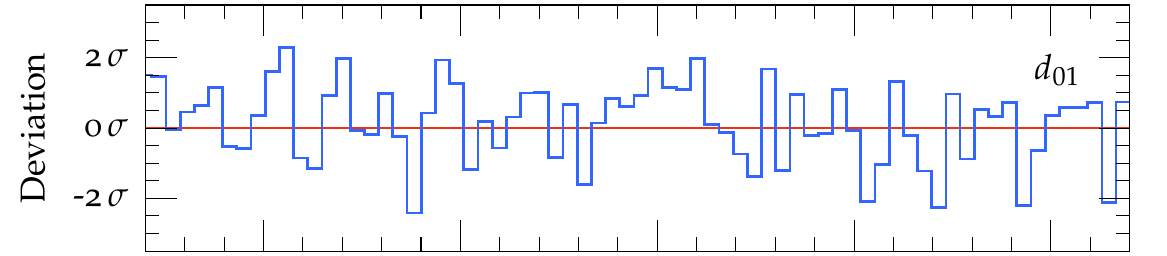}\\[-1mm]
    \includegraphics[width=\textwidth]{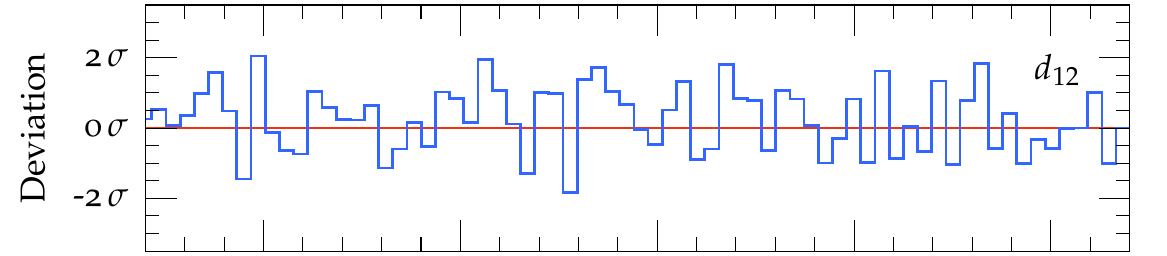}\\[-1mm]
    \includegraphics[width=\textwidth]{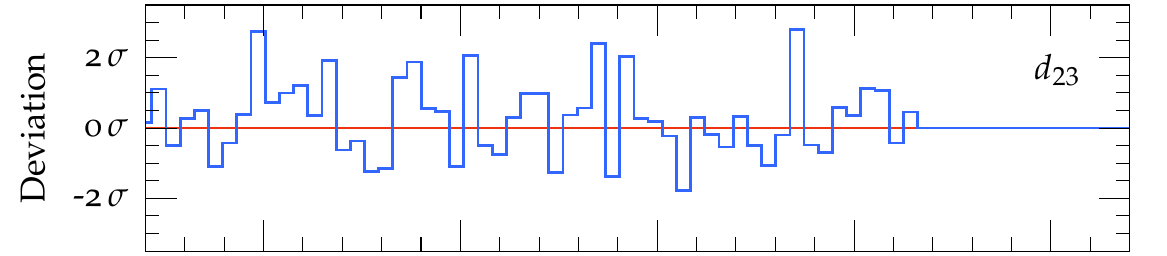}\\[-1mm]
    \includegraphics[width=\textwidth]{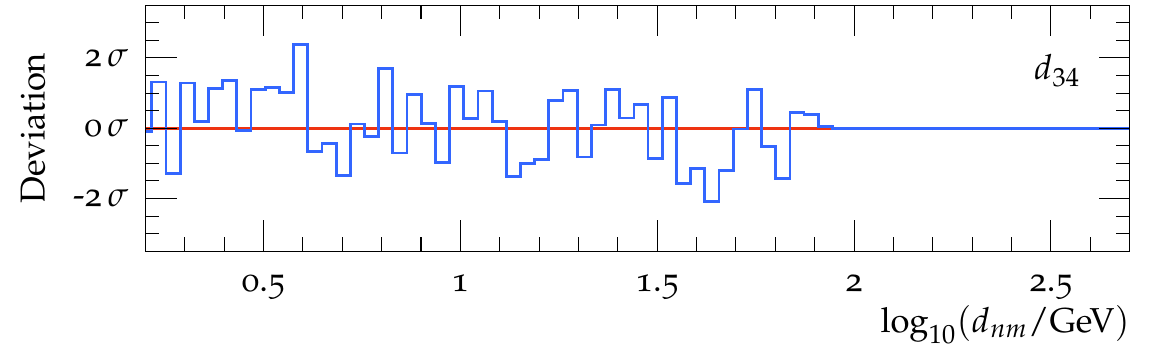}
  \end{minipage}\hskip 5mm
  \caption{Validation in $q\bar{q}\to e^+e^-$ and $gg\to\tau^+\tau^-$
    \label{fig:validation_pp}}
\end{figure}

Figures~\ref{fig:validation_ee},~\ref{fig:validation_ep} and~\ref{fig:validation_pp} each
show a comparison between the results from \Dire{}+\Pythia and \Dire{}+\Sherpa. 
Each simulation contains $10^8$ events. The lower panels present the deviation between the two 
predictions, normalized to the statistical uncertainty of \Dire{}+\Sherpa in the respective bin. 
This distribution should exhibit statistical fluctuations only. We validate quark
splitting functions in the reactions $e^+e^-\to\text{hadrons}$ (Fig.~\ref{fig:validation_ee} left),
$e^+p\to e^+\text{jet}$ (Fig.~\ref{fig:validation_ep} left), and $pp\to e^+e^-$ 
(Fig.~\ref{fig:validation_pp} left). These three cases cover all possible dipole types,
i.e.\ final-state splitter with final-state spectator, final-state splitter with initial-state
spectator and vice versa, and initial-state splitter with initial-state spectator.
Gluon splitting functions are validated in the reactions $\tau^+\tau^-\to\text{hadrons}$ 
(Fig.~\ref{fig:validation_ee} right), $\tau^+p\to \tau^+\text{jet}$ (Fig.~\ref{fig:validation_ep} 
right), and $pp\to\tau^+\tau^-$ (Fig.~\ref{fig:validation_pp} left), all mediated by Higgs-boson
exchange using HEFT~\cite{Ellis:1975ap,*Wilczek:1977zn,*Shifman:1979eb,*Ellis:1979jy}.
\clearpage

\section{Results}
\label{sec:results}
In this section we compare \Dire predictions from \Sherpa~\cite{Gleisberg:2003xi,*Gleisberg:2008ta}
with experimental data. When applicable, we use the CT10nlo PDF set~\cite{Lai:2010vv}
and the corresponding strong coupling. We employ the kinematics scheme from App.~\ref{sec:css_kin_if}.
Our results include the simulation of QED radiation
in the case of Drell-Yan lepton pair production~\cite{Schonherr:2008av}, and hadronization
in the case of $e^+e^-\to$ hadrons~\cite{Winter:2003tt}. Otherwise they are given at the 
parton level in order to exhibit the features of the \Dire shower only.
Analyses are performed with \Rivet~\cite{Buckley:2010ar}.

\begin{figure}[p]
  \centering
  \includegraphics[width=6cm]{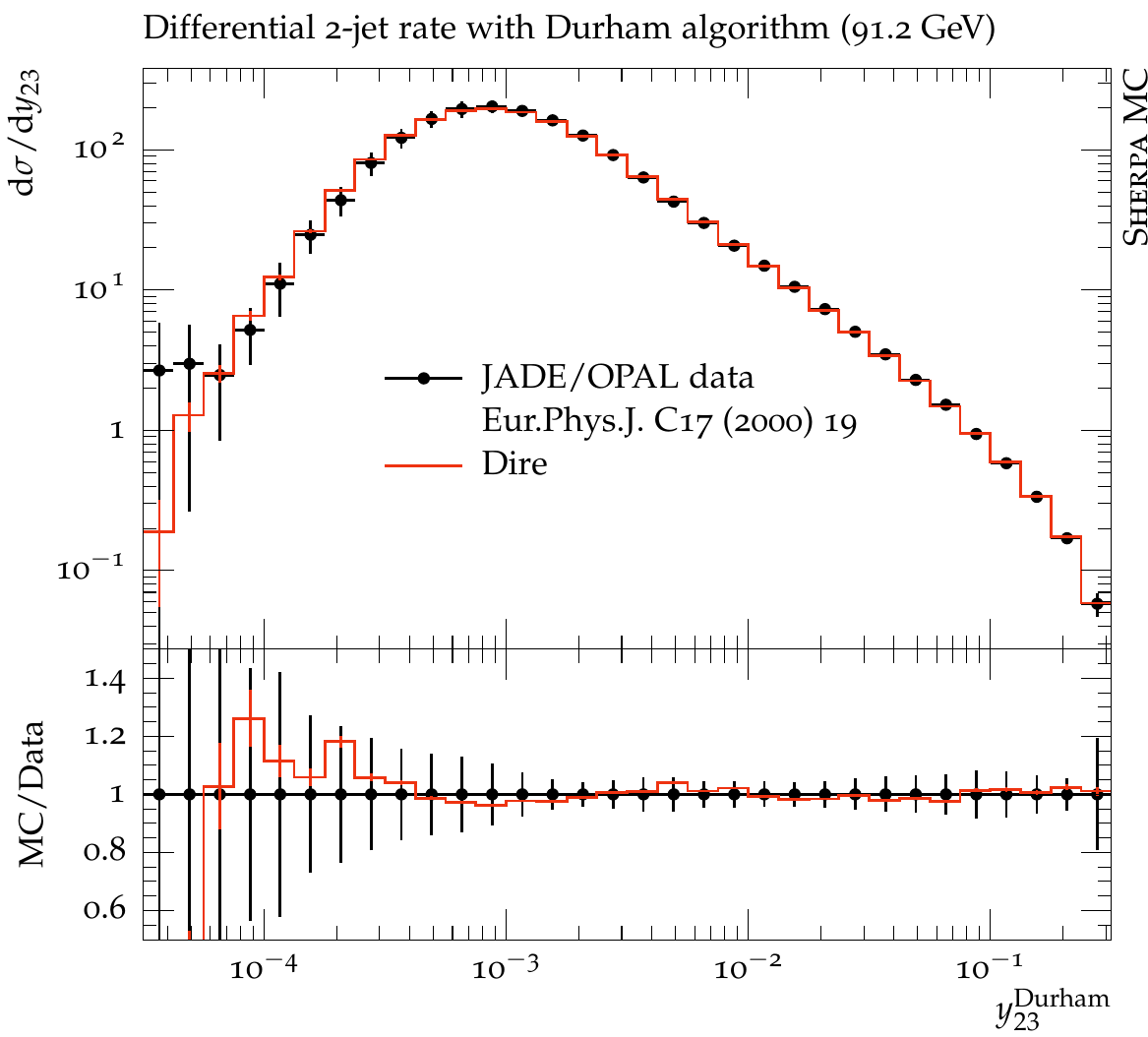}
  \includegraphics[width=6cm]{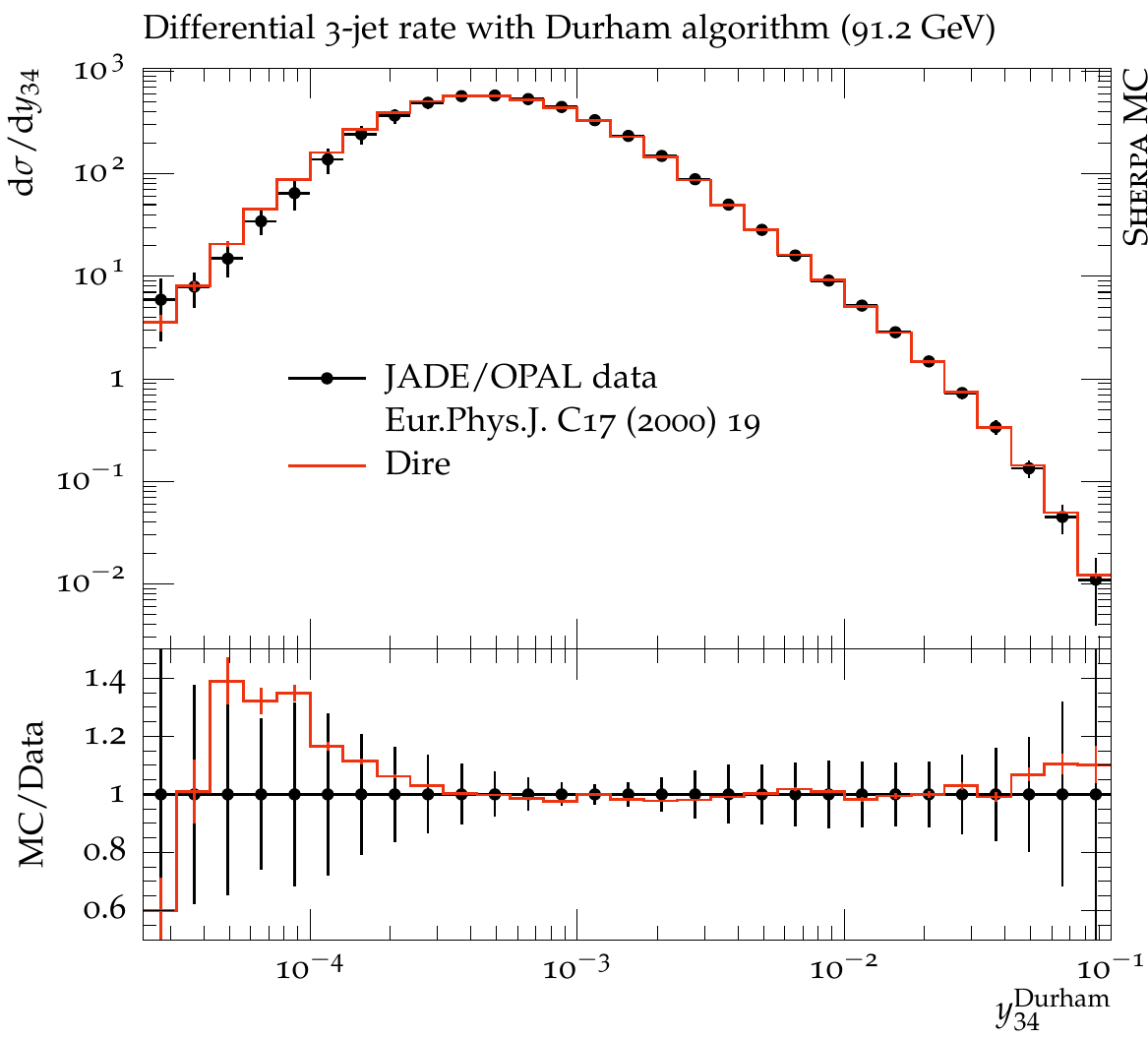}\\
  \includegraphics[width=6cm]{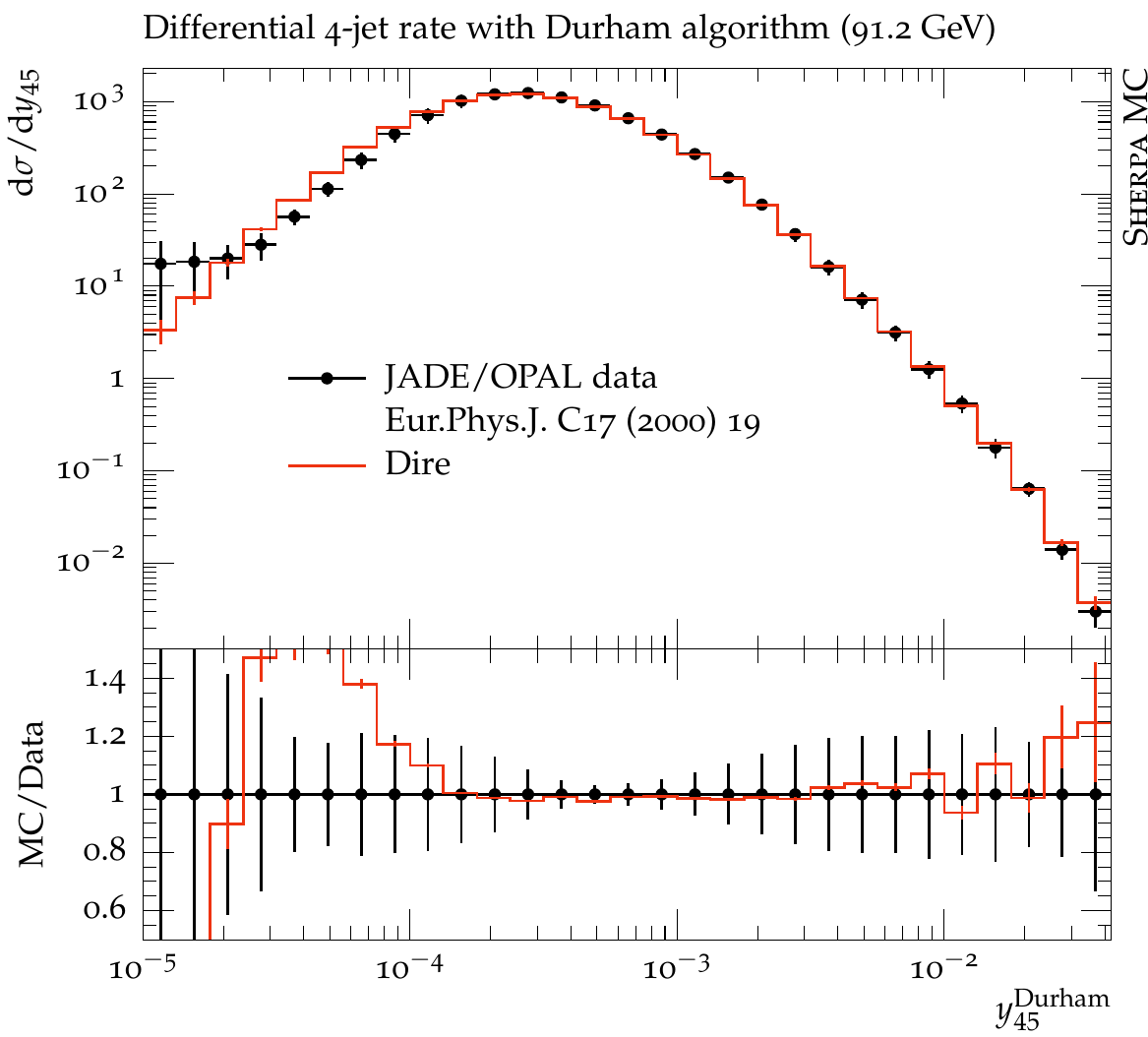}
  \includegraphics[width=6cm]{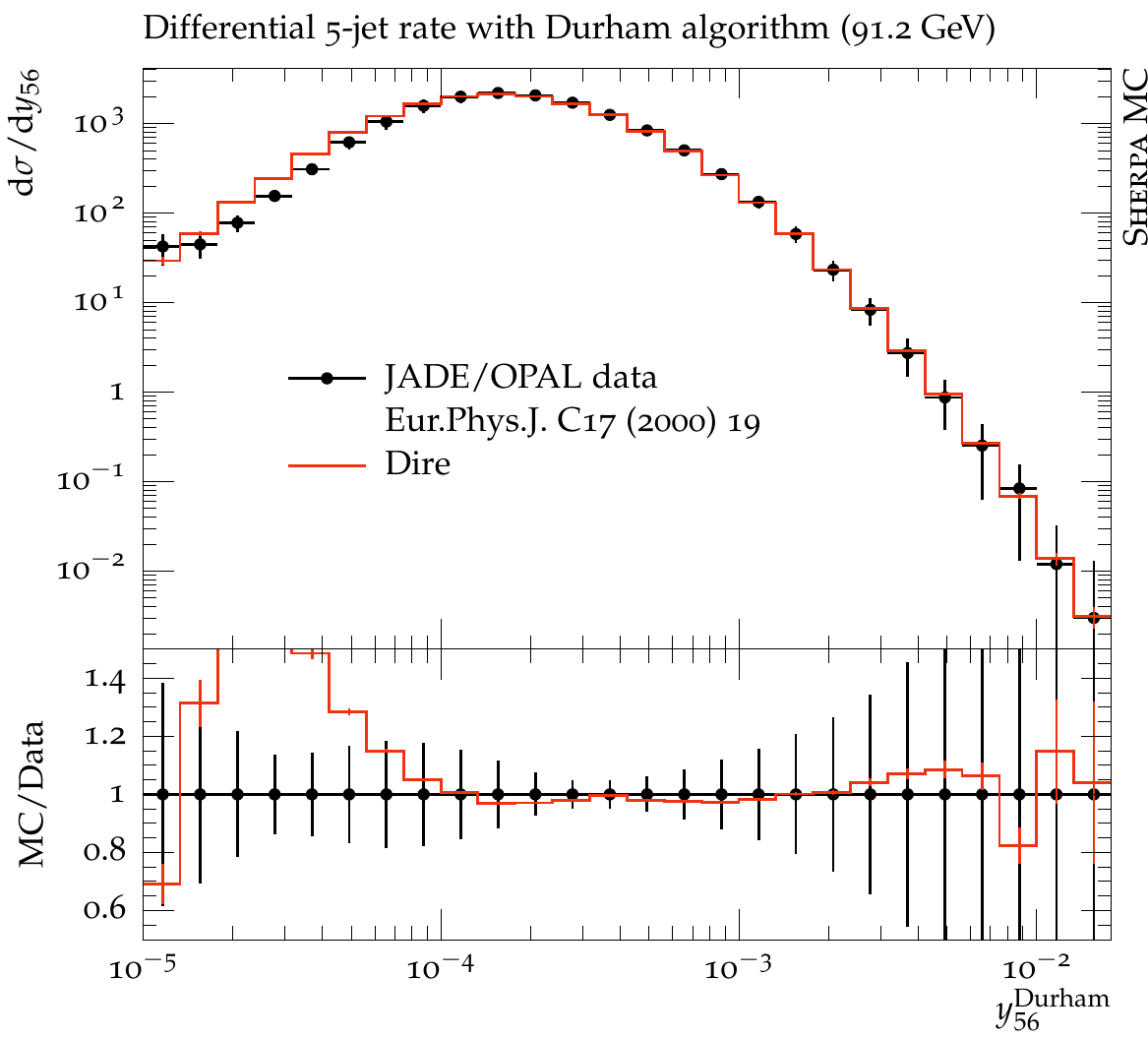}
  \caption{\Dire predictions in comparison to LEP data from~\cite{Pfeifenschneider:1999rz}.
    \label{fig:lep_jetrates}}
\end{figure}

Figure~\ref{fig:lep_jetrates} shows predictions from the \Dire parton shower for differential 
jet rates in the Durham scheme compared to experimental results from the JADE and OPAL 
collaborations~\cite{Pfeifenschneider:1999rz}. The perturbative region is to the right
of the plot, and $y\sim 2.8\cdot10^{-3}$ corresponds to the $b$-quark mass. The simulation
of nonperturbative effects dominates the predictions below $\sim10^{-4}$. We observe that,
in the perturbative region, the results are in excellent agreement with the experimental
measurements.

\begin{figure}[p]
  \centering
  \includegraphics[width=6cm]{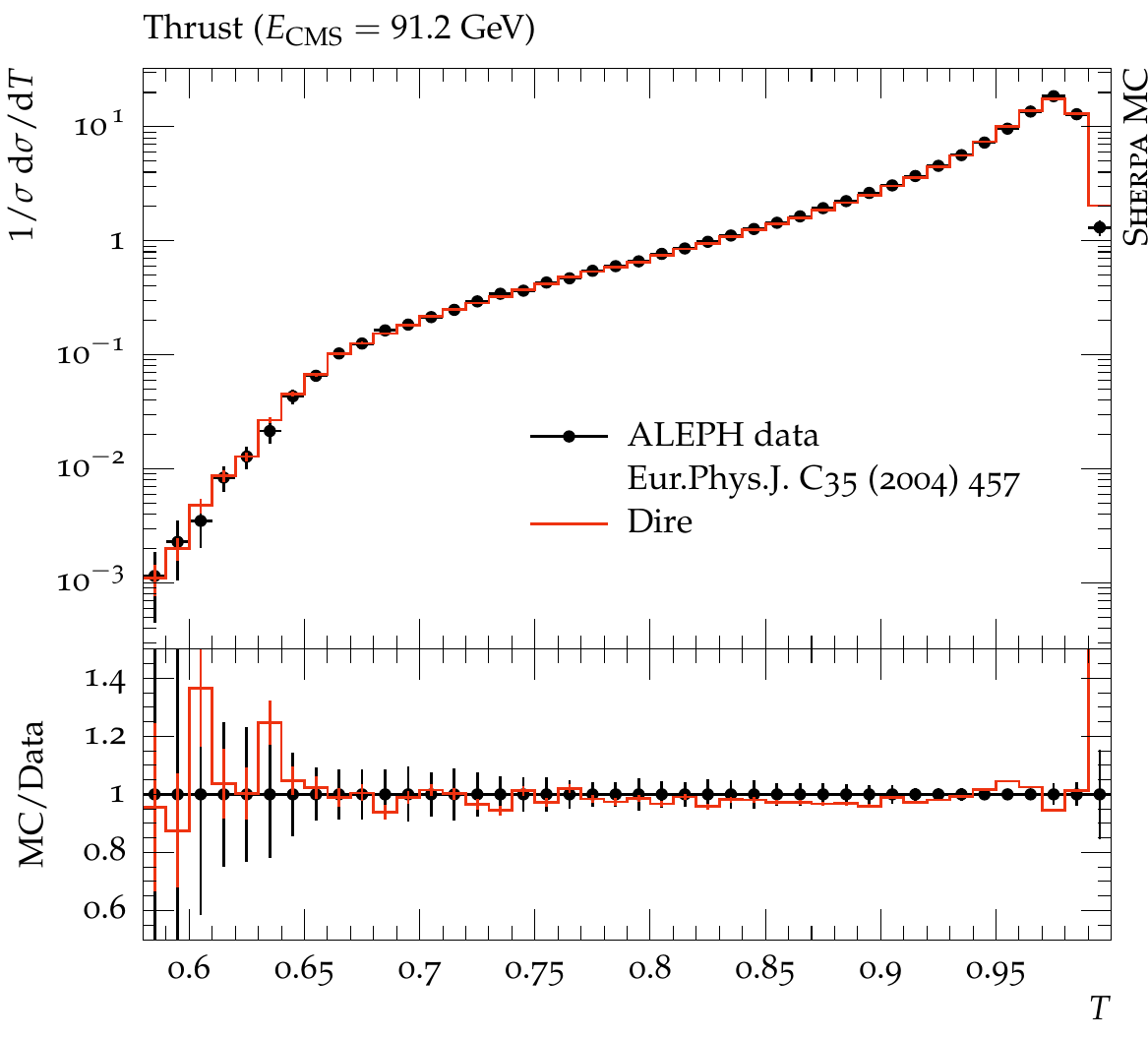}
  \includegraphics[width=6cm]{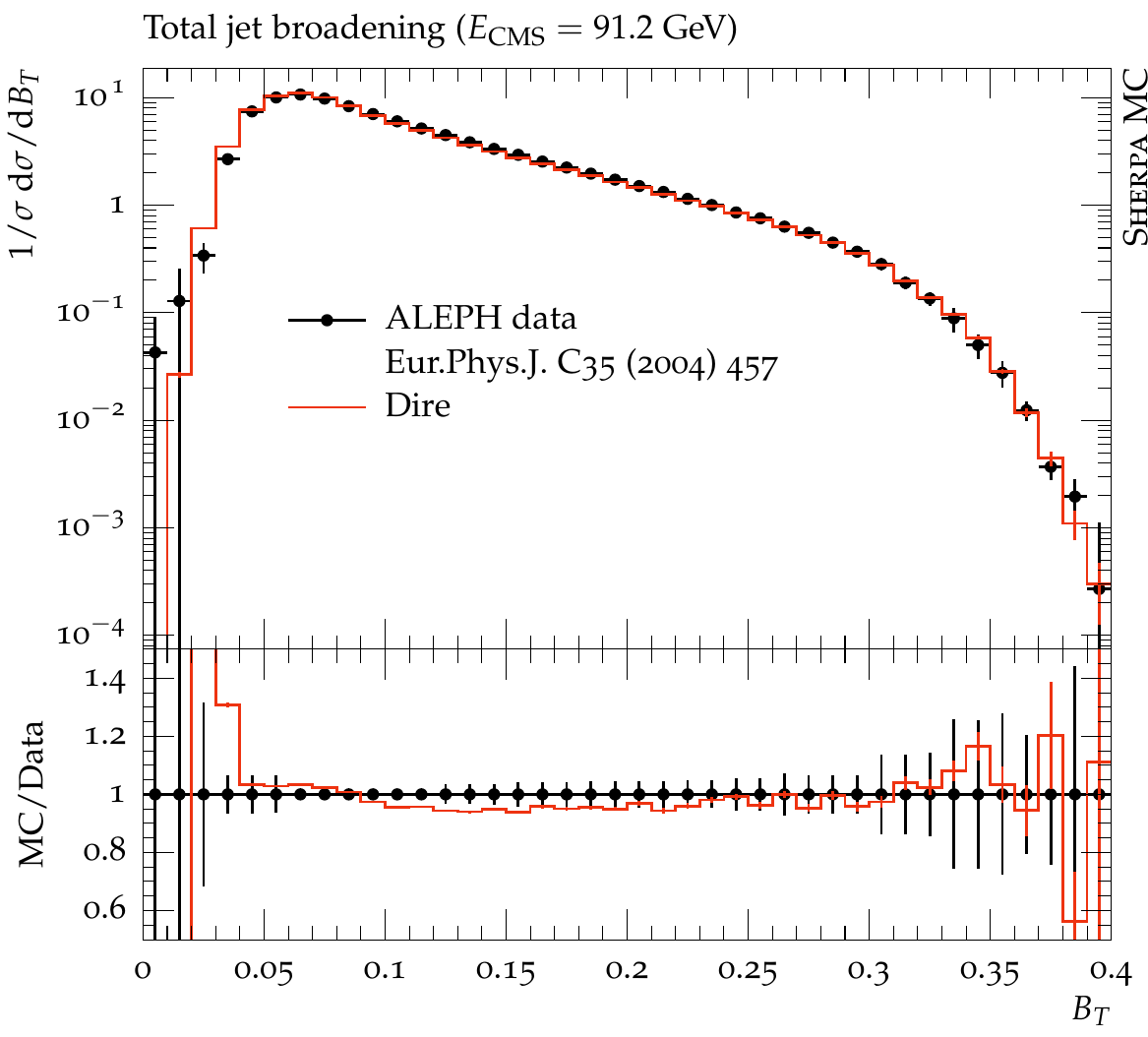}
  \includegraphics[width=6cm]{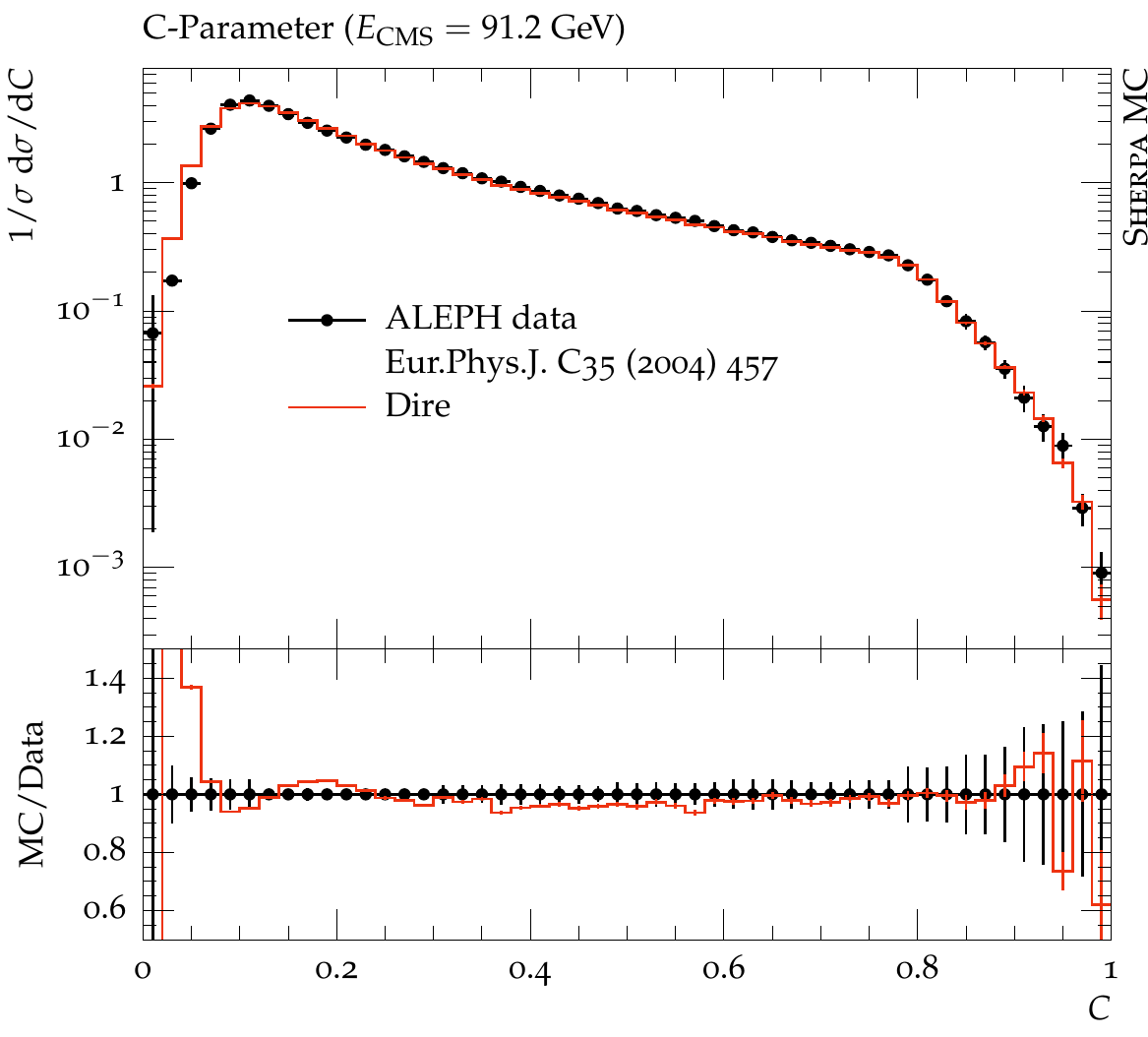}
  \includegraphics[width=6cm]{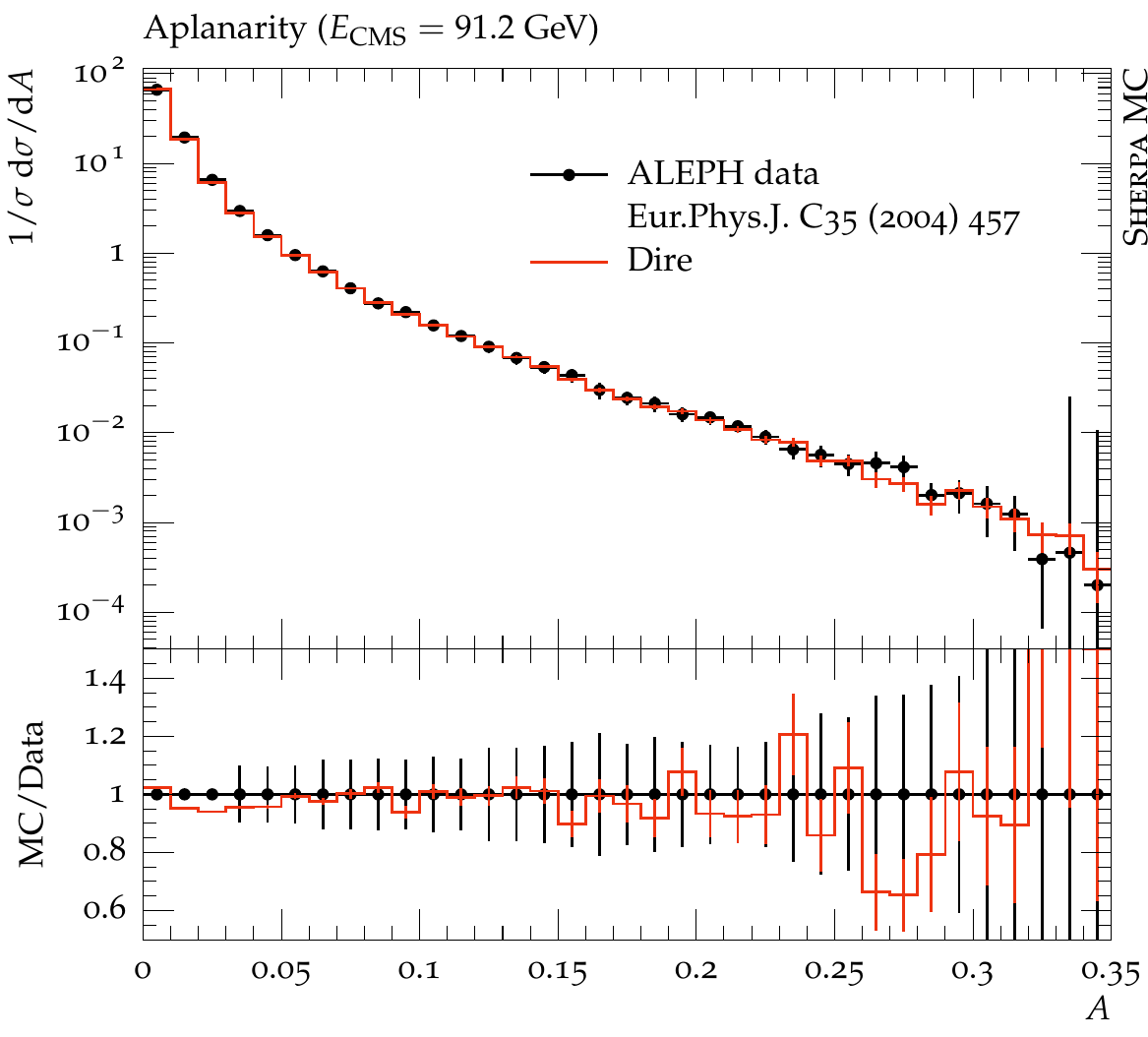}
  \caption{\Dire predictions in comparison to LEP data from~\cite{Heister:2003aj}.
    \label{fig:lep_shapes}}
\end{figure}

Figure~\ref{fig:lep_shapes} shows a comparison for event shapes measured by the
ALEPH collaboration~\cite{Heister:2003aj}. The perturbative region is to the right
of the plot, except for the thrust distribution, where it is to the left. We notice
some deviation in the predictions for jet broadening and for the $C$-parameter. 
However, these deviations are mostly within the 2$\sigma$ uncertainty of the
experimental measurements, and they occur close to the nonperturbative region.
It can also be expected that the simulations improve upon including matrix-element
corrections or when merging the \Dire shower with higher-multiplicity calculations. 
This has been demonstrated, for example, in~\cite{Lavesson:2008ah,*Gehrmann:2012yg}.

\begin{figure}[p]
  \centering
  \includegraphics[width=7.5cm]{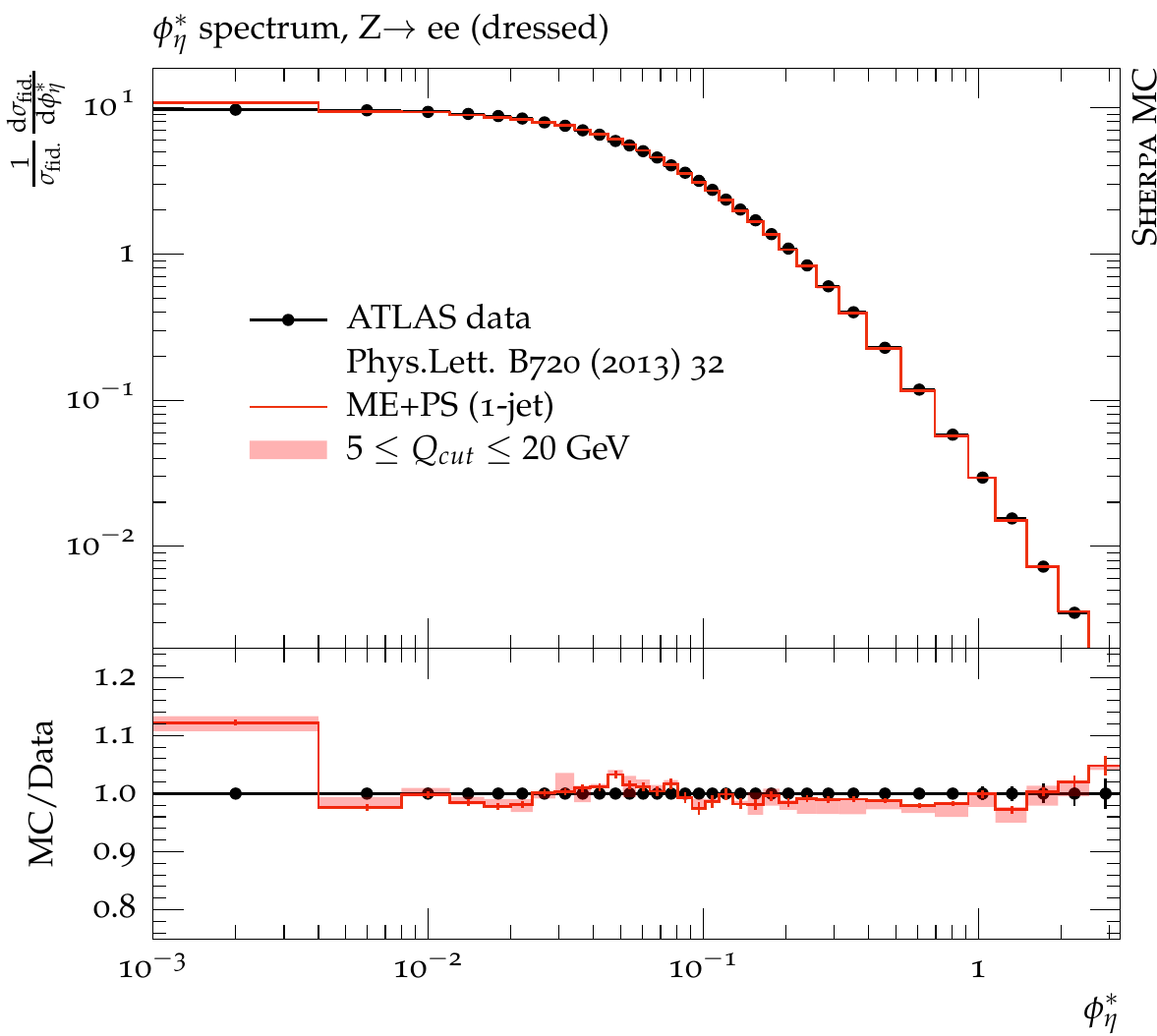}
  \includegraphics[width=7.5cm]{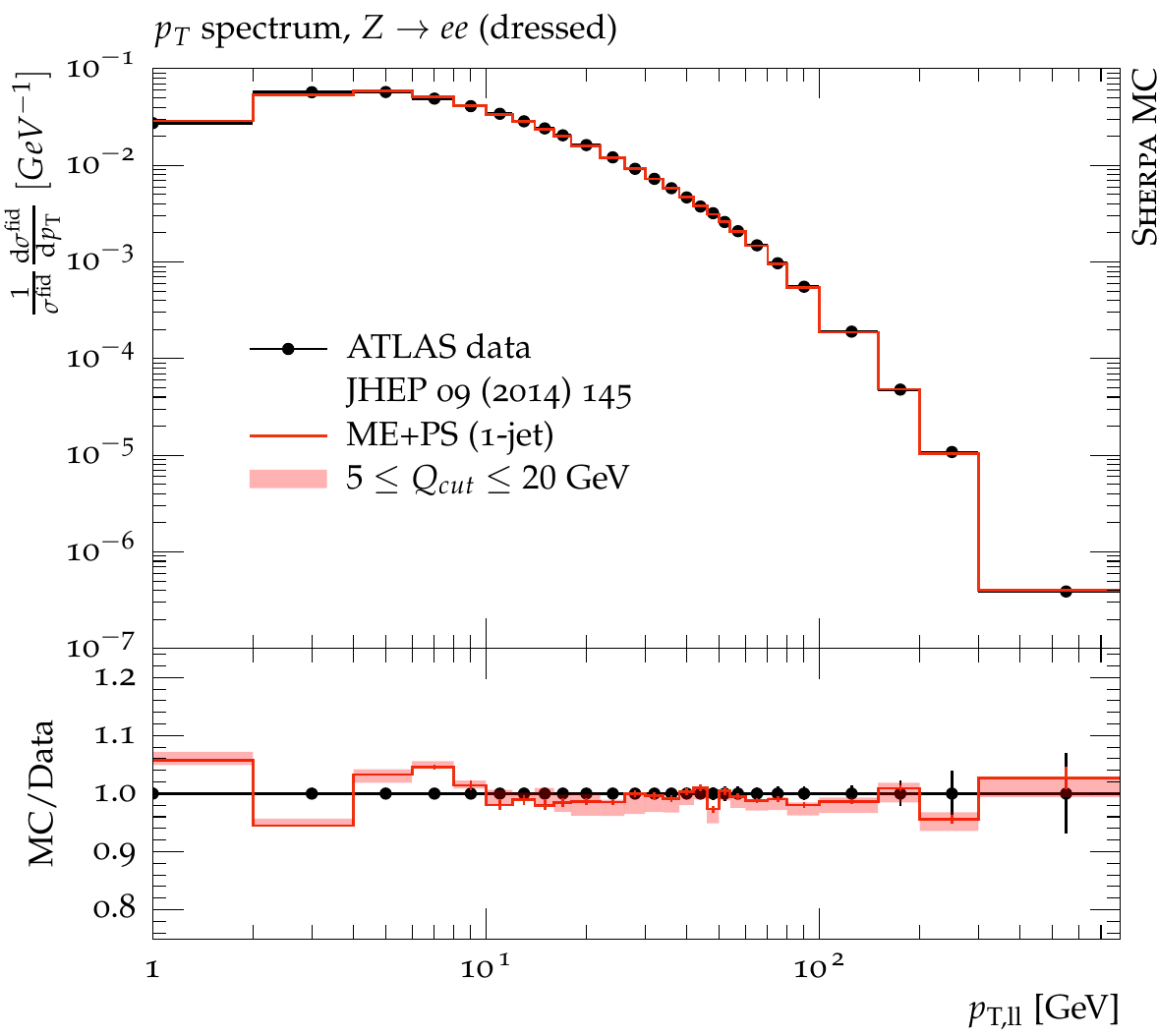}\\
  \begin{minipage}{7.5cm}
    \includegraphics[width=\textwidth]{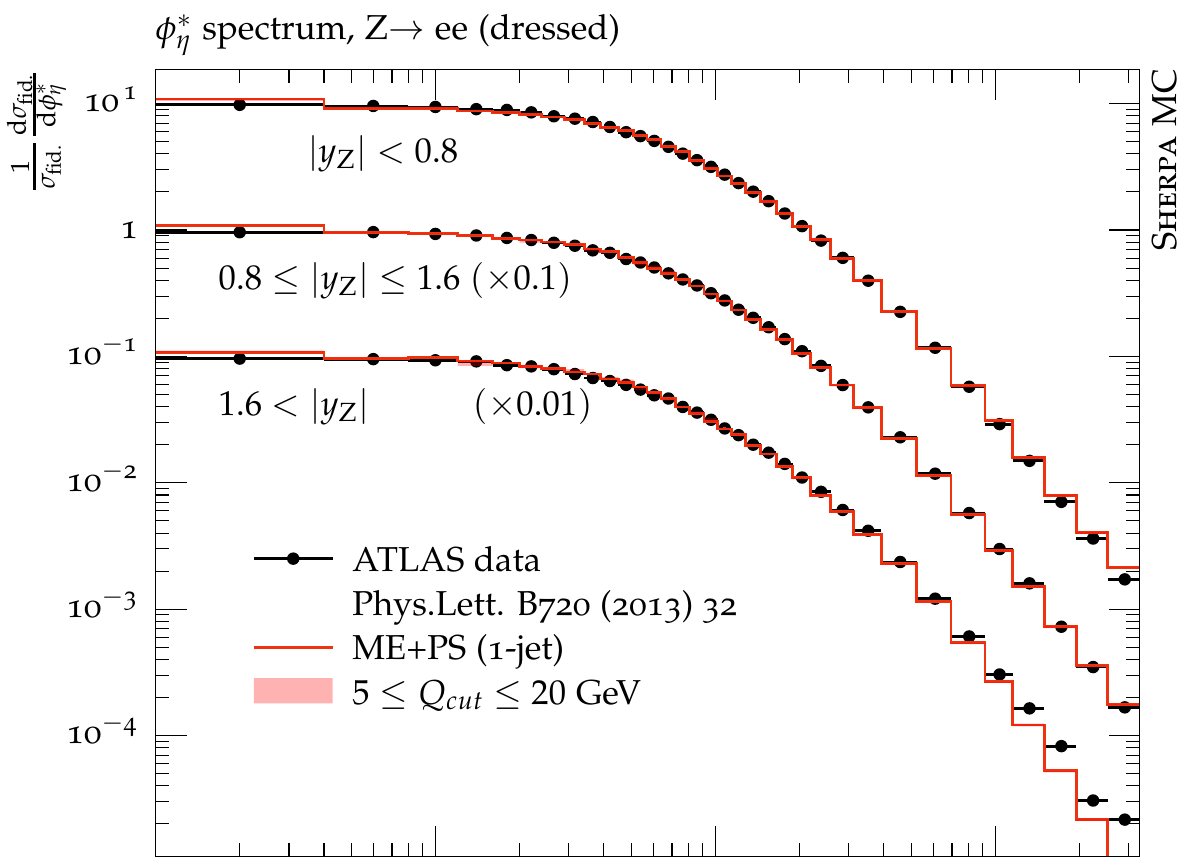}\\[-1mm]
    \includegraphics[width=\textwidth]{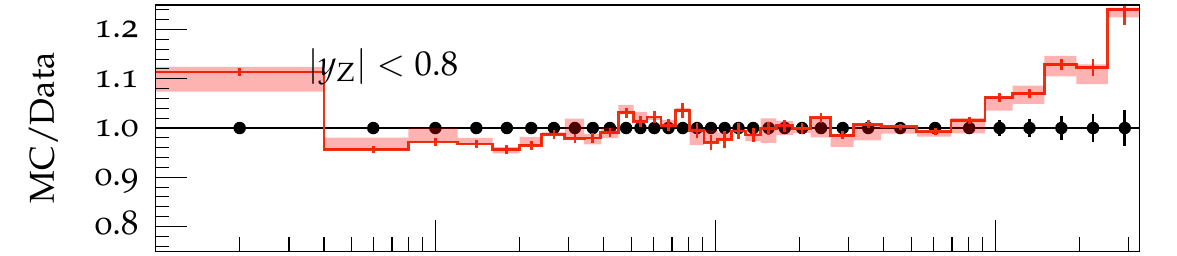}\\[-1mm]
    \includegraphics[width=\textwidth]{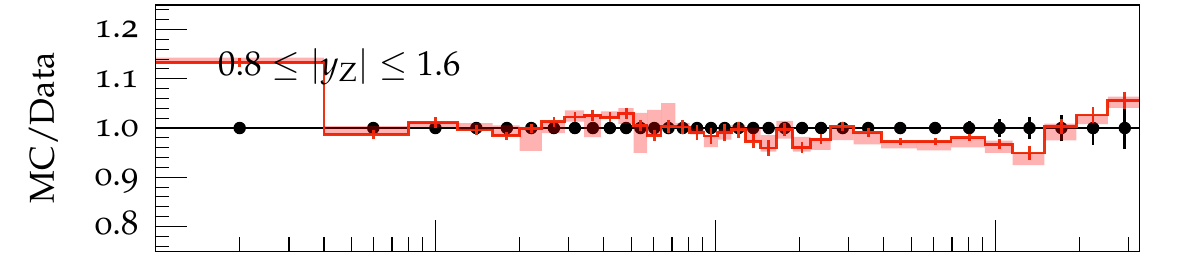}\\[-1mm]
    \includegraphics[width=\textwidth]{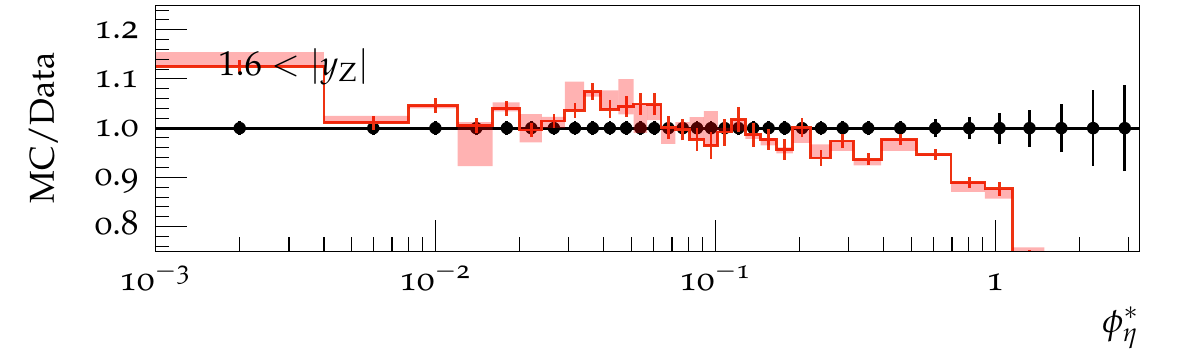}
  \end{minipage}
  \begin{minipage}{7.5cm}
    \includegraphics[width=\textwidth]{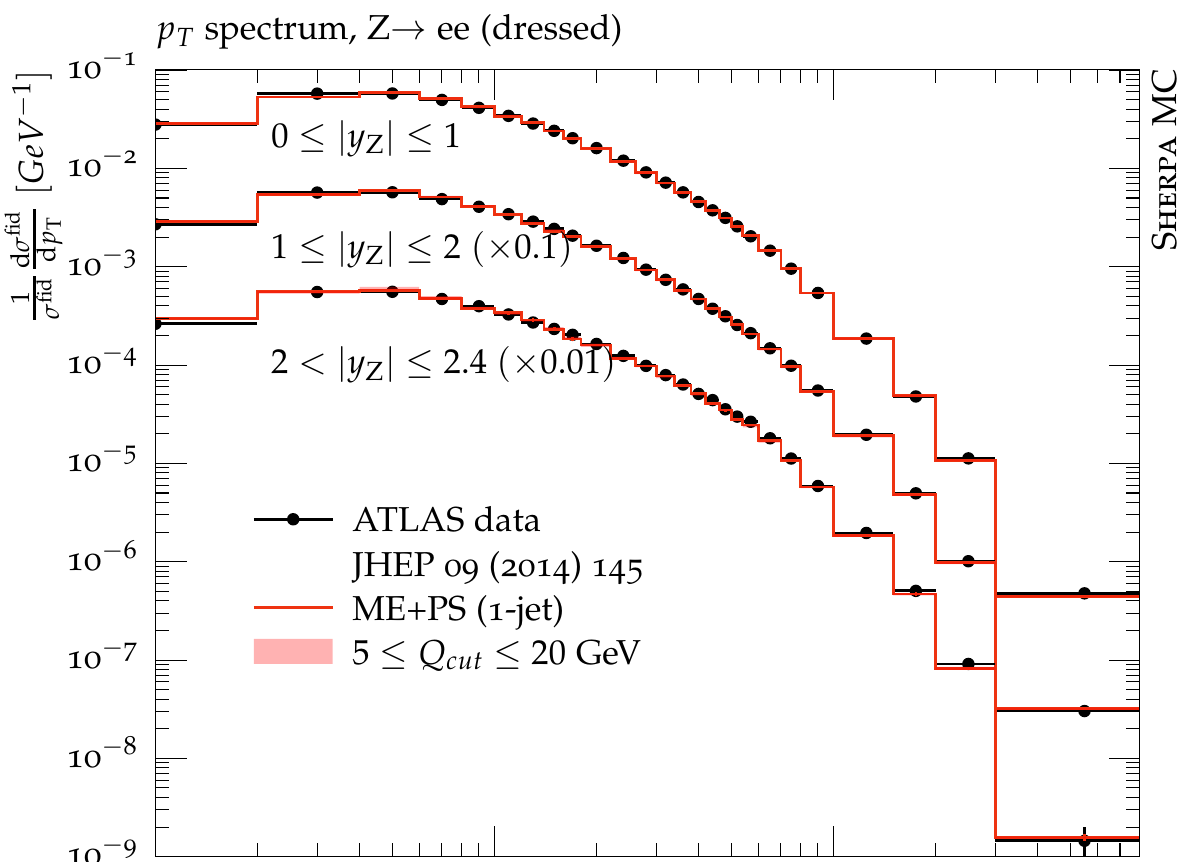}\\[-1mm]
    \includegraphics[width=\textwidth]{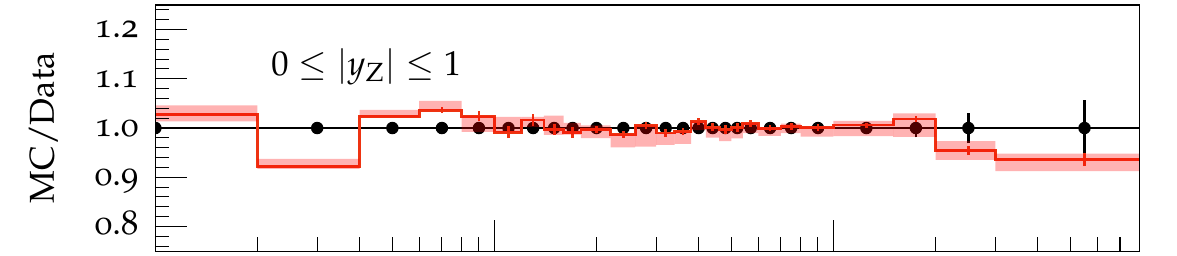}\\[-1mm]
    \includegraphics[width=\textwidth]{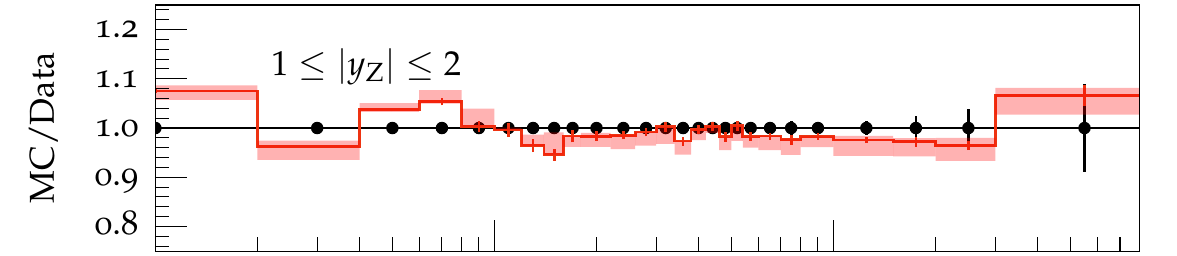}\\[-1mm]
    \includegraphics[width=\textwidth]{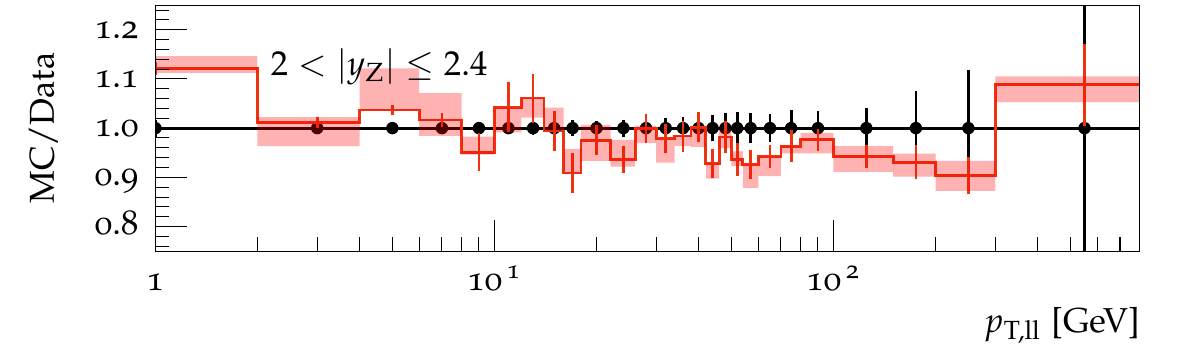}
  \end{minipage}
  \caption{\Dire ME+PS merged predictions in comparison to ATLAS data from~\cite{Aad:2012wfa} 
    and~\cite{Aad:2014xaa}.
    \label{fig:lhc_ptspectra}}
\end{figure}

Figure~\ref{fig:lhc_ptspectra} shows angular correlations in comparison to ATLAS data 
from~\cite{Aad:2012wfa}, and the transverse momentum spectrum of the Drell-Yan lepton pair 
in comparison to ATLAS data from~\cite{Aad:2014xaa}. It is well known that pure parton shower
predictions are insufficient to describe these measurements. Therefore, we merge our parton shower 
with 1-jet matrix elements using the CKKW-L procedure~\cite{Catani:2001cc}.
In order to assess the related uncertainties, we vary the merging cut by a factor 2 around 
the central value of $Q_{\rm cut}=10~{\rm GeV}$. The associated uncertainty band is shown 
in light red. The size of the variation is comparable to the statistical uncertainties,
which are displayed as error bars on the Monte-Carlo prediction.

\begin{figure}[p]
  \centering
  \begin{minipage}{7.5cm}
    \includegraphics[width=\textwidth]{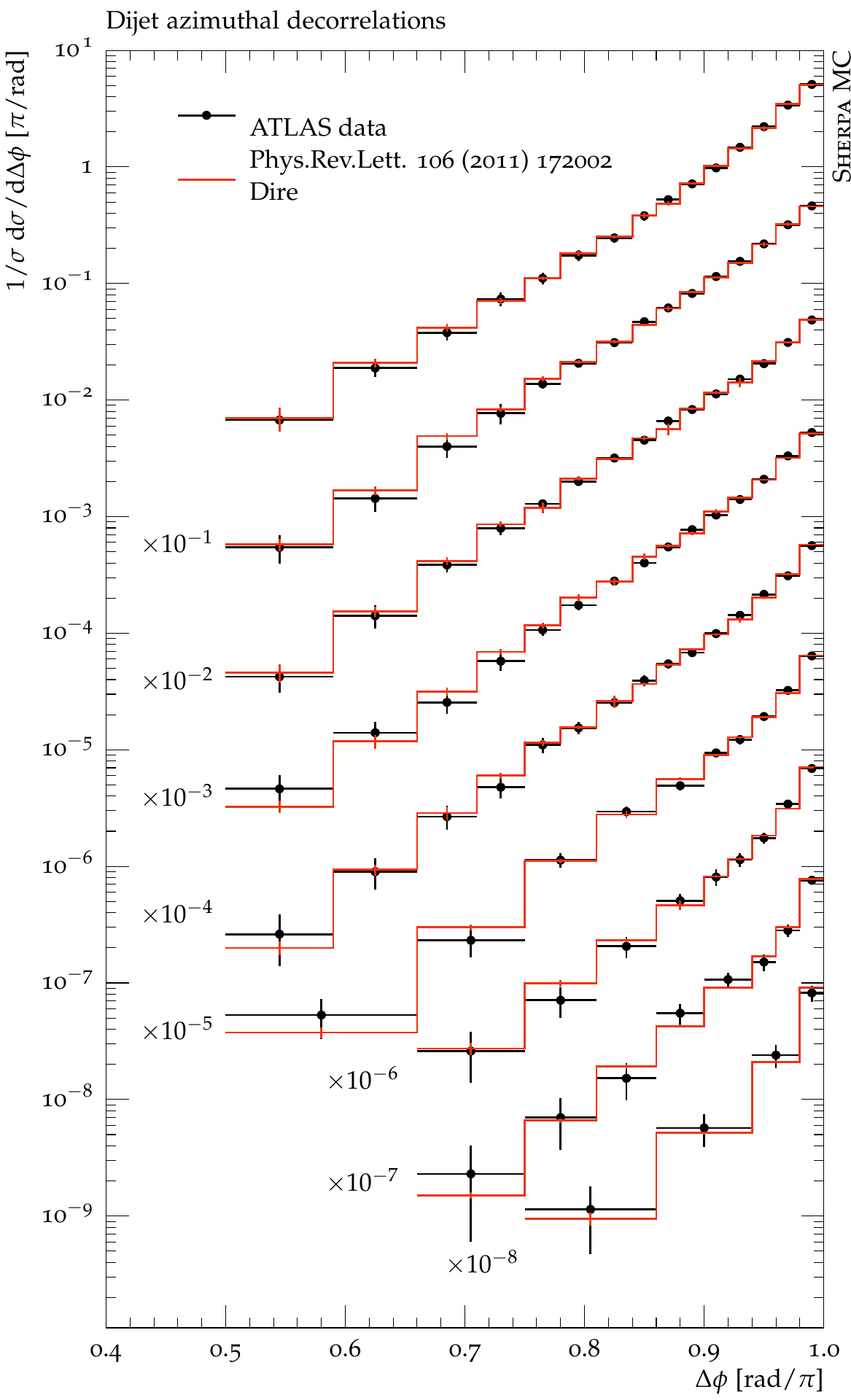}
  \end{minipage}
  \begin{minipage}{7.5cm}\vskip 4mm
    \includegraphics[width=\textwidth]{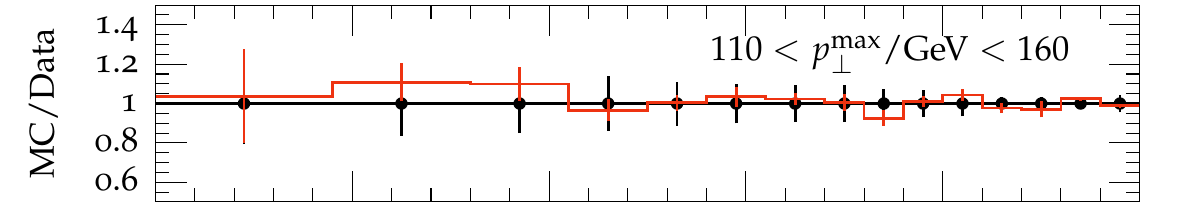}\\[-1mm]
    \includegraphics[width=\textwidth]{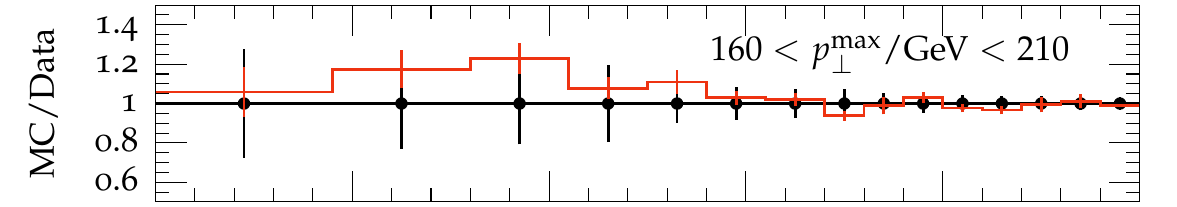}\\[-1mm]
    \includegraphics[width=\textwidth]{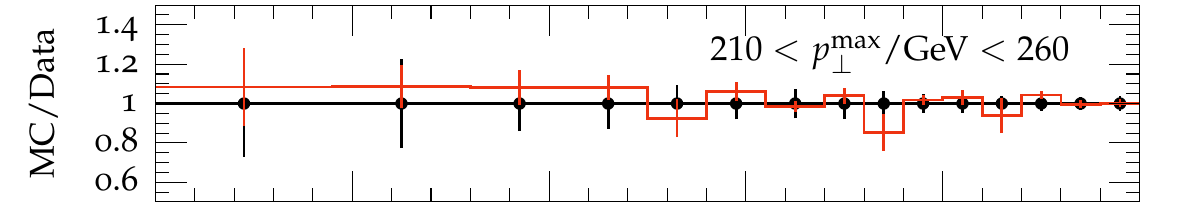}\\[-1mm]
    \includegraphics[width=\textwidth]{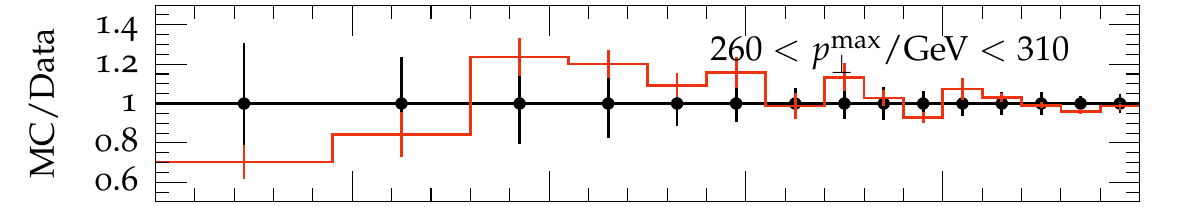}\\[-1mm]
    \includegraphics[width=\textwidth]{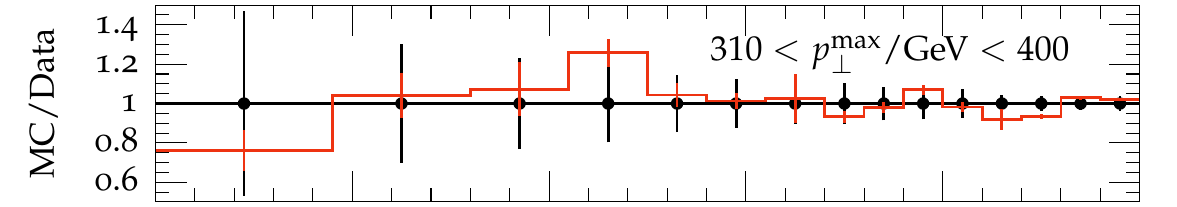}\\[-1mm]
    \includegraphics[width=\textwidth]{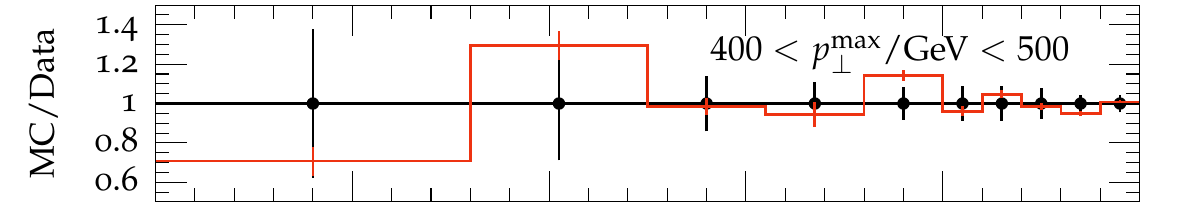}\\[-1mm]
    \includegraphics[width=\textwidth]{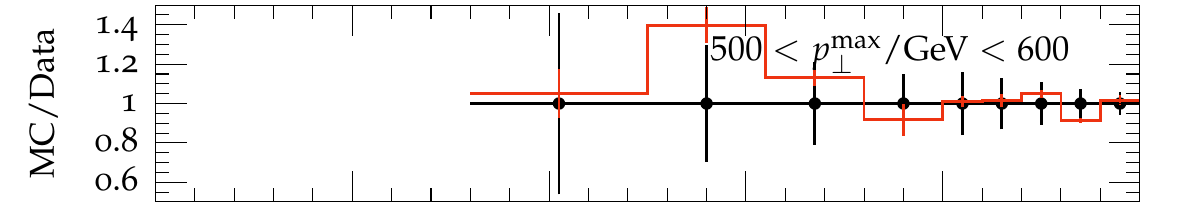}\\[-1mm]
    \includegraphics[width=\textwidth]{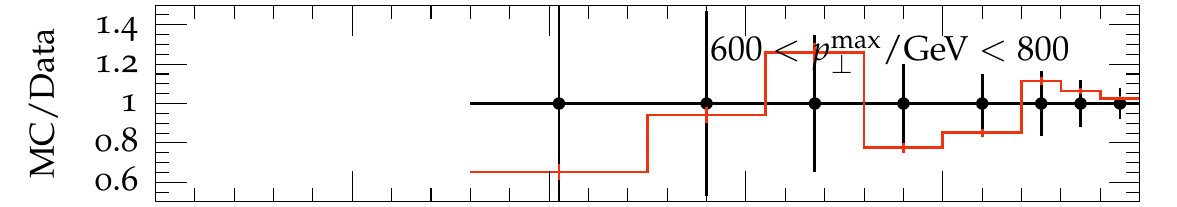}\\[-1mm]
    \includegraphics[width=\textwidth]{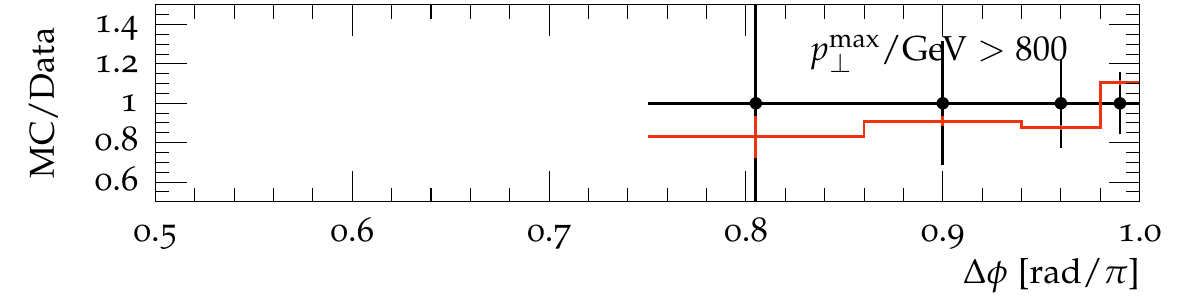}
  \end{minipage}
  \caption{\Dire predictions in comparison to ATLAS data from~\cite{daCosta:2011ni}.
    \label{fig:lhc_decorrelations}}
\end{figure}

Figure~\ref{fig:lhc_decorrelations} shows di-jet azimuthal decorrelations in different regions
of jet transverse momentum. We compare \Dire predictions with experimental results from the 
ATLAS collaboration~\cite{daCosta:2011ni}. This observable tests for higher-order effects 
in some detail~\cite{Wobisch:2015jea}.

\section{Conclusions}
\label{sec:conclusions}
We presented a new dipole-like parton-shower algorithm, that is constructed along very simple
arguments: Firstly, the ordering variable should exhibit a symmetry in emitter and spectator
momenta, such that the dipole-like picture can be re-interpreted as a dipole picture in the
soft limit. At the same time, the splitting functions are regularized in the soft anti-collinear 
region using partial fractioning of the soft eikonal in the Catani-Seymour approach. 
They are then modified to satisfy the ordinary sum rules in the collinear limit. This leads
to an invariant formulation of the parton-shower algorithm, which is in complete analogy to the
standard DGLAP case. We computed the anomalous dimensions, which match previous results for dipole-like
parton showers. We presented first phenomenologically relevant predictions using the new algorithm,
and we observe very good agreement with experimental data from LEP and LHC experiments.

\section*{Acknowledgments}
We thank Frank Krauss, Ye Li, Leif L{\"o}nnblad, Marek Sch{\"o}nherr and Torbj{\"o}rn Sj{\"o}strand
for many enlightening discussions and for their comments on the manuscript.
We are grateful to Valerio Bertone and Juan Rojo for pointing out an error in the definition
of the splitting functions.
This work was supported by the US Department of Energy under contract DE--AC02--76SF00515.


\clearpage

\appendix
\section{Parton-shower kinematics}
\label{sec:kinematics}
The precise algorithm for constructing the splitting kinematics depends on the type
of splitter and spectator parton. There are four separate cases. Note that initial-state
partons are assumed to be massless for collinear PDF evolution to be valid. In practically
implemented parton-shower algorithms they are often taken massive instead, in order to
obtain a better description of experimental data. Therefore we give the kinematics formulae
with full mass-dependence, including initial-state parton masses.\footnote{We use the 
  notation and definitions of~\cite{Catani:1996vz}, both in the massive and in the massless case. 
  This implies $\tilde{z}_i=(p_ip_k)/(p_ip_k+p_jp_k)$ and $y_{ij,k}=2p_ip_j/Q^2$ for final-final dipoles,
  $\tilde{z}_i=(p_ip_a)/(p_ip_a+p_jp_a)$ and $2\,x_{ij,a}=Q^2/(p_ip_a+p_jp_a)$ for final-initial dipoles,
  $\tilde{u}_j=p_jp_a/(p_jp_a+p_kp_a)$ and $2\,x_{jk,a}=Q^2/(p_jp_a+p_kp_a)$ for initial-final dipoles, and
  $\tilde{x}_{j,ab}=Q^2/(2p_ap_b)$ and $\tilde{v}_{j}=(p_ap_j)/(p_ap_b)$ for initial-initial dipoles.}

\subsection{Final-state splitter with final-state spectator}
\label{sec:css_kin_ff}
\begin{enumerate}
\item Determine the new momentum of the spectator parton as
  \begin{equation}\label{eq:def_ff_pk}
    \begin{split}
      p_k^{\,\mu}=&\;\left(\tilde{p}_k^{\,\mu}-
        \frac{q\cdot\tilde{p}_k}{q^2}\,q^\mu\right)\,
      \sqrt{\frac{\lambda(q^2,s_{ij},m_k^2)}{\lambda(q^2,m_{ij}^2,m_k^2)}}
      +\frac{q^2+m_k^2-s_{ij}}{2\,q^2}\,q^\mu\;,
    \end{split}
  \end{equation}
  with $\lambda$ denoting the K{\"a}llen function 
  $\lambda(a,b,c)=\rbr{a-b-c}^2-4\,bc$\\ and
  $s_{ij}\,=\;y_{ij,k}\,(q^2-m_k^2)+(1-y_{ij,k})\,(m_i^2+m_j^2)$.
\item Construct the new momentum of the emitter parton, $p_i$, as
  \begin{align}\label{eq:def_ff_pi_pj}
    p_i^\mu\,=&\;\bar{z}_i\,\frac{\gamma(q^2,s_{ij},m_k^2)\,p_{ij}^\mu
      -s_{ij}\,p_k^\mu}{\beta(q^2,s_{ij},m_k^2)}
    +\frac{m_i^2+{\rm k}_\perp^2}{\bar{z}_i}\,
    \frac{p_k^\mu-m_k^2/\gamma(q^2,s_{ij},m_k^2)\,p_{ij}^\mu}{
      \beta(q^2,s_{ij},m_k^2)}+k_\perp^\mu\;,
  \end{align}
  where $q=\tilde{p}_{ij}+\tilde{p}_k$, 
  $\beta(a,b,c)={\rm sgn}\rbr{a-b-c}\sqrt{\lambda(a,b,c)}$,\\
  $2\,\gamma(a,b,c)=\rbr{a-b-c}+\beta(a,b,c)$ and $p_{ij}^\mu=q^\mu-p_k^\mu$.\\
  The parameters $\bar{z}_i$ and ${\rm k}_\perp^2=-k_\perp^2$ 
  of this decomposition are given by
  \begin{equation}\label{eq:def_ff_zi_kt}
    \begin{split}
      \bar{z}_i\,=&\;\frac{q^2-s_{ij}-m_k^2}{\beta\rbr{q^2,s_{ij},m_k^2}}\,
      \sbr{\;\tilde{z}_i\,-\,\frac{m_k^2}{\gamma(q^2,s_{ij},m_k^2)}
          \frac{s_{ij}+m_i^2-m_j^2}{q^2-s_{ij}-m_k^2}}\;,\\
      {\rm k}_\perp^2\,=&\;\bar{z}_i\,(1-\bar{z}_i)\,s_{ij}-
        (1-\bar{z}_i)\, m_i^2-\bar{z}_i\, m_j^2\;,
    \end{split}
  \end{equation}
\end{enumerate}
In the massless case, this algorithm reduces to the following~\cite{Schumann:2007mg}
\begin{equation}
  \begin{split}
    p_i^\mu=&\;\tilde{z}_i\,\tilde{p}_{ij}^\mu+y_{ij,k}(1-\tilde{z}_i)\,\tilde{p}_k^\mu+k_\perp^\mu\;,\\
    p_j^\mu=&\;(1-\tilde{z}_i)\,\tilde{p}_{ij}^\mu+y_{ij,k}\tilde{z}_i\,\tilde{p}_k^\mu-k_\perp^\mu\;,\\
    p_k^\mu=&\;(1-y_{ij,k})\,\tilde{p}_k^\mu\;,
  \end{split}
\end{equation}
where $\tilde{z}_i=(z-y_{ij,k})/(1-y_{ij,k})$
and $y_{ij,k}=\kappa^2/(1-z)$, cf.\ Sec.~\ref{sec:dire}.

\subsection{Final-state splitter with initial-state spectator}
\label{sec:css_kin_fi}
\begin{enumerate}
\item Determine the new momentum of the spectator parton as
  \begin{equation}\label{eq:def_fi_pa}
    \begin{split}
      p_a^{\,\mu}=&\;\left(\tilde{p}_a^{\,\mu}-\frac{q\cdot\tilde{p}_a}{q_\parallel^2}\,
      q_\parallel^{\,\mu}\right)\,
      \sqrt{\frac{\lambda(q^2,s_{ij},m_a^2)-4\,m_a^2\,\tilde{p}_{ij\perp}^2}{
              \lambda(q^2,m_{ij}^2,m_a^2)-4\,m_a^2\,\tilde{p}_{ij\perp}^2}}
      +\frac{q^2+m_a^2-s_{ij}}{2\,q_\parallel^2}\,q_\parallel^{\,\mu}\;,
    \end{split}
  \end{equation}
  where $q=\tilde{p}_a-\tilde{p}_{ij}$, $q_\parallel=q+\tilde{p}_{ij\perp}$
  and $s_{ij}=(1-1/x_{ij,a})\,(q^2-m_a^2)+(m_i^2+m_j^2)/x_{ij,a}$.
\item Proceed as in Sec.~\ref{sec:css_kin_ff}, except that 
  $m_k\to m_a$ and $p_k\to -p_a$.
\end{enumerate}
In the massless case, this algorithm reduces to the following~\cite{Schumann:2007mg}
\begin{equation}
  \begin{split}
    p_i^\mu=&\;\tilde{z}_i\,\tilde{p}_{ij}^\mu+(1-\tilde{z}_i)\,\frac{1-x_{ij,a}}{x_{ij,a}}\,\tilde{p}_a^\mu+k_\perp^\mu\;,\\
    p_j^\mu=&\;(1-\tilde{z}_i)\,\tilde{p}_{ij}^\mu+\tilde{z}_i\,\frac{1-x_{ij,a}}{x_{ij,a}}\,\tilde{p}_a^\mu-k_\perp^\mu\;,\\
    p_a^\mu=&\;\frac{1}{x_{ij,a}}\,\tilde{p}_a^\mu\;,
  \end{split}
\end{equation}
where $\tilde{z}_i=z$ and $x_{ij,a}=1-\kappa^2/(1-z)$, cf.\ Sec.~\ref{sec:dire}.

\subsection{Initial-state splitter with final-state spectator (local recoil)}
\label{sec:css_kin_if}
\begin{enumerate}
\item Determine the new momentum of the splitting parton as
  \begin{equation}\label{eq:def_if_pa_2}
    \begin{split}
      p_a^{\,\mu}=&\;\left(\tilde{p}_{aj}^{\,\mu}-\frac{q\cdot\tilde{p}_{aj}}{
        q_\parallel^2}\,q_\parallel^{\,\mu}\right)\,
      \sqrt{\frac{\lambda(q^2,s_{jk},m_a^2)-4\,m_a^2\,\tilde{p}_{k\perp}^2}{
              \lambda(q^2,m_k^2,m_{aj}^2)-4\,m_{aj}^2\,\tilde{p}_{k\perp}^2}}
      +\frac{q^2+m_a^2-s_{jk}}{2\,q_\parallel^2}\,q_\parallel^{\,\mu}\;,
    \end{split}
  \end{equation}
  where $q=\tilde{p}_{aj}-\tilde{p}_k$, $q_\parallel=q+\tilde{p}_{k\perp}$
  and $s_{jk}=(1-1/x_{jk,a})\,(q^2-m_a^2)+(m_j^2+m_k^2)/x_{jk,a}$.
\item Proceed as in Sec.~\ref{sec:css_kin_ff}, except that $\tilde{z}_j\to u_j$,
  $m_k\to m_a$, $m_j\to m_k$, $p_k\to -p_a$ and $p_j\to p_k$.
\end{enumerate}
In the massless case, this algorithm reduces to the following~\cite{Schumann:2007mg}
\begin{equation}
  \begin{split}
    p_a^\mu=&\;\frac{1}{x_{jk,a}}\,\tilde{p}_{aj}^\mu\;,\\
    p_j^\mu=&\;(1-u_j)\,\frac{1-x_{jk,a}}{x_{jk,a}}\,\tilde{p}_{aj}^\mu+u_j\,\tilde{p}_k^\mu-k_\perp^\mu\;,\\
    p_k^\mu=&\;u_j\,\frac{1-x_{jk,a}}{x_{jk,a}}\,\tilde{p}_{aj}^\mu+(1-u_j)\,\tilde{p}_k^\mu+k_\perp^\mu\;,
  \end{split}
\end{equation}
where $x_{jk,a}=z$ and $u_j=\kappa^2/(1-z)$, cf.\ Sec.~\ref{sec:dire}.

\subsection{Initial-state splitter with final-state spectator (global recoil)}
\label{sec:css_kin_if_res}
This scheme can be chosen as an alternative to the one in Sec.~\ref{sec:css_kin_if}.
The algorithm is equivalent to the method outlined in~\cite{Hoeche:2009xc,Carli:2009cg}.
It alleviates a formal problem with transverse momentum resummation in Drell-Yan type processes~\cite{Nagy:2009vg},
but it is numerically less stable, and therefore employed only if $\abs{x_{jka}-u_j}>\varepsilon$, where $\varepsilon\sim 10^{-4}$.
\begin{enumerate}
\item Determine the new momentum of the spectator parton as
  \begin{equation}\label{eq:def_if_pa_1}
    \begin{split}
      p_k^{\,\mu}=&\;\left(\tilde{p}_k^{\,\mu}-\frac{q\cdot\tilde{p}_k}{q^2}\,
        q^{\,\mu}\right)\,
      \sqrt{\frac{\lambda(q^2,s_{aj},m_k^2)}{\lambda(q^2,m_{aj}^2,m_k^2)}}
      +\frac{q^2+m_k^2-s_{aj}}{2\,q^2}\,q^{\,\mu}\;,
    \end{split}
  \end{equation}
  where $q=\tilde{p}_k-\tilde{p}_{aj}$ and
  $s_{aj}=u_j/x_{jk,a}\,(q^2-m_k^2)+(m_a^2+m_j^2)\,(1-u_j)/x_{jk,a}$.
\item Construct the momentum of the emitted particle, $p_j$, as
  \begin{align}\label{eq:def_if_pi_pj}
    p_j^{\,\mu}\,=&\;-\bar{z}_j\,\frac{\gamma(q^2,s_{aj},m_k^2)\,
      p_{aj}^{\,\mu}+s_{aj}\,p_k^{\,\mu}}{\beta(q^2,s_{aj},m_k^2)}+
      \frac{m_j^2+{\rm k}_\perp^2}{\bar{z}_j}\,\frac{
      p_k^{\,\mu}+m_k^2/\gamma(q^2,s_{aj},m_k^2)\,p_{aj}^{\,\mu}}{
      \beta(q^2,s_{aj},m_k^2)}+k_\perp^\mu\;,
  \end{align}
  The parameters $\bar{z}_j$ and ${\rm k}_\perp^2=-k_\perp^2$ 
  of this decomposition are given by
  \begin{equation}\label{eq:def_if_zi_kt}
    \begin{split}
      \bar{z}_j\,=&\;\frac{q^2-s_{aj}-m_k^2}{\beta(q^2,s_{aj}^2,m_k^2)}\,
      \sbr{\;\frac{x_{jk,a}-1}{x_{jk,a}-u_i}\,-\,
          \frac{m_k^2}{\gamma(q^2,s_{aj},m_k^2)}
          \frac{s_{aj}+m_j^2-m_a^2}{q^2-s_{aj}-m_k^2}}\;,\\
      {\rm k}_\perp^2\,=&\;\bar{z}_j\,(1-\bar{z}_j)\,s_{aj}
        -(1-\bar{z}_j)\,m_j^2-\bar{z}_j\, m_a^2\;,
    \end{split}
  \end{equation}
\item Boost $p_a$ and all final state particles into the frame where
  $p_{a}$ is aligned along the beam direction,\\
  with $p_b$, the opposite-side beam particle, unchanged.
\end{enumerate}
In the massless case, this algorithm reduces to the following~\cite{Platzer:2009jq}
\footnote{We changed the definition of the transverse momentum
  in Eqs.~\eqref{eq:if_global_ml} compared to Ref.~\cite{Platzer:2009jq},
  in order to match Eq.~\eqref{eq:def_if_zi_kt}.}
\begin{equation}\label{eq:if_global_ml}
  \begin{split}
    p_a^\mu=&\;\frac{1-u_j}{x_{jk,a}-u_j}\,\tilde{p}_{aj}^\mu
      +\frac{u_j}{x_{jk,a}}\frac{1-x_{jk,a}}{x_{jk,a}-u_j}\,\tilde{p}_k^\mu
      +k_\perp^\mu\;,\\
    p_j^\mu=&\;\frac{1-x_{jk,a}}{x_{jk,a}-u_j}\,\tilde{p}_{aj}^\mu
      +\frac{u_j}{x_{ij,a}}\frac{1-u_j}{x_{jk,a}-u_j}\,\tilde{p}_k^\mu
      +k_\perp^\mu\;,\\
    p_k^\mu=&\Big(1-\frac{u_j}{x_{jk,a}}\Big)\,\tilde{p}_k^\mu\;.
  \end{split}
\end{equation}
Note that, following the arguments in~\cite{Platzer:2009jq}, the light-cone 
momentum fraction entering the PDF is still given by $x/z=x/x_{ij,k}$.

\subsection{Initial-state splitter with initial-state spectator}
\label{sec:css_kin_ii}
\begin{enumerate}
\item Determine the new momentum of the splitting parton as
  \begin{equation}\label{eq:def_ii_pk}
    \begin{split}
      p_a^{\,\mu}=&\;\left(\tilde{p}_a^{\,\mu}-\frac{\tilde{m}_{aj}^2}{
        \gamma(q^2,\tilde{m}_{aj}^2,m_b^2)}\,p_b^{\,\mu}\right)\,
      \sqrt{\frac{\lambda(s_{ab},m_a^2,m_b^2)}{
        \lambda(q^2,\tilde{m}_{aj}^2,m_b^2)}}
      +\frac{m_a^2}{\gamma(s_{ab},m_a^2,m_b^2)}\,p_b^{\,\mu}\;,
    \end{split}
  \end{equation}
  where $q=\tilde{p}_a+p_b$ and
  $s_{ab}=(q^2-m_j^2)/x_{j,ab}+(1-1/x_{j,ab})\,(m_a^2+m_b^2)$.
\item Construct the momentum of the emitted parton, $p_j$, as
  \begin{align}\label{eq:def_ii_pi_pj}
    p_j^{\,\mu}\,=&\;(1-\bar{z}_{aj})\,\frac{
      \gamma(s_{ab},m_a^2,m_b^2)\,p_a^{\,\mu}-m_a^2\,p_b^{\,\mu}}{
      \beta(s_{ab},m_a^2,m_b^2)}+
      \frac{m_j^2+{\rm k}_\perp^2}{1-\bar{z}_{aj}}\,\frac{
      p_b^{\,\mu}-m_b^2/\gamma(s_{ab},m_a^2,m_b^2)\,p_a^{\,\mu}}{
      \beta(s_{ab},m_a^2,m_b^2)}-k_\perp^\mu\;,
  \end{align}
  The parameters $\bar{z}_{aj}$ and ${\rm k}_\perp^2=-k_\perp^2$ 
  of this decomposition are given by
  \begin{equation}\label{eq:def_ii_zi_kt}
    \begin{split}
      \bar{z}_{aj}\,=&\;\frac{s_{ab}-m_a^2-m_b^2}{\beta(s_{ab},m_a^2,m_b^2)}\,
      \sbr{\;(x_{j,ab}+v_j)\,-\,\frac{m_b^2}{\gamma(s_{ab},m_a^2,m_b^2)}
          \frac{s_{aj}+m_a^2-m_j^2}{s_{ab}-m_a^2-m_b^2}}\;,\\
      {\rm k}_\perp^2\,=&\;\bar{z}_{aj}\,(1-\bar{z}_{aj})\,m_a^2
        -(1-\bar{z}_{aj})\,s_{aj}-\bar{z}_{aj}\, m_j^2\;,
    \end{split}
  \end{equation}
\item Boost all remaining final-state particles into the frame defined by $p_a+p_b-p_j$,
  using the algorithm defined in Sec.~(5.5) of~\cite{Catani:1996vz}.
\end{enumerate}
In the massless case, this algorithm reduces to the following~\cite{Schumann:2007mg}
\begin{equation}
  \begin{split}
    p_a^\mu=&\;\frac{1}{x_{j,ab}}\,\tilde{p}_{aj}^\mu\;,\\
    p_j^\mu=&\;\frac{1-x_{j,ab}-\tilde{v}_j}{x_{j,ab}}\,\tilde{p}_{aj}^\mu+\tilde{v}_j\,\tilde{p}_b^\mu-k_\perp^\mu\;,\\
    k_i^\mu=&\;\Lambda(\tilde{p}_{ai}+p_b,p_a+p_b-p_ i)^\mu_{\;\nu}\,\tilde{k}_i^\nu\;,
  \end{split}
\end{equation}
where $x_{j,ab}=z-\tilde{v}_j$ and $\tilde{v}_j=\kappa^2/(1-z)$, cf.\ Sec.~\ref{sec:dire},
and where $k_i$ stands for any final-state momentum, including EW particles.
The Lorentz transformation, $\Lambda$, is given by
\begin{equation}
  \Lambda(\tilde{K},K)^\mu_{\;\nu}=g^\mu_{\;\nu}
  -\frac{2\,(K+\tilde{K})^\mu(K+\tilde{K})_\nu}{(K+\tilde{K})^2}
  +\frac{2\,K^\mu\tilde{K}_\nu}{\tilde{K}^2}\;,
\end{equation}

The inverted kinematics needed for matching the \Dire shower to NLO computations
and for merging with higher-order tree-level calculations are given in~\cite{Catani:2002hc}.
The combined momenta are obtained from Eqs.~\eqref{eq:def_ff_pk}, \eqref{eq:def_fi_pa},
\eqref{eq:def_if_pa_2} and~\eqref{eq:def_ii_pk} by swapping momenta and masses
before and after emission. In the case of final-state emitter with final-state spectator,
for example, this amounts to the replacement $p_k\leftrightarrow \tilde{p}_k$ and
$s_{ij}\leftrightarrow m_{ij}^2$.

\section{Anomalous dimensions}
\label{sec:anom_dim}
The anomalous dimensions of the splitting functions listed in Eqs.~\eqref{eq:ap_kernels}
are computed in this appendix as
\begin{equation}
  \gamma_{ab}(N,\kappa^2)=\int_0^1\done z\, z^N P_{ab}(z,\kappa^2)\;.
\end{equation}
They are given by
\begin{equation}
  \begin{split}
    \gamma_{qq}(N,\kappa^2)=&\;2C_F\,\Gamma(N,\kappa^2)
    -\frac{C_F\,(2N+3)}{(N+1)(N+2)}+\gamma_q\\
    \gamma_{gq}(N,\kappa^2)=&\;2C_F\,{\rm K}(N,\kappa^2)-\frac{C_F\,(N+3)}{(N+1)(N+2)}\\
    \gamma_{gg}(N,\kappa^2)=&\;2C_A\,\Gamma(N,\kappa^2)
    +2C_A\,{\rm K}(N,\kappa^2)-\frac{2C_A\,(N+3)}{(N+1)(N+2)}-\frac{2C_A}{N+3}
    +\gamma_g\\
    \gamma_{qg}(N,\kappa^2)=&\;-\frac{T_R\,N}{(N+1)(N+2)}+\frac{2T_R}{N+3}
  \end{split}
\end{equation}
where
\begin{equation}\label{eq:gamma_soft_css}
  \begin{split}
    2\Gamma(N,\kappa^2)=&\;
    \frac{2\,_3F_2\left(1,1,\frac{3}{2};\frac{N+3}{2},\frac{N+4}{2};
      -\kappa^{-2}\right)}{(N+1)(N+2)\,\kappa^2}
    -\ln\frac{1+\kappa^2}{\kappa^2}\;,\\
    2{\rm K}(N,\kappa^2)=&\;
    \frac{2\,_2F_1\left(1,\frac{N+2}{2};\frac{N+4}{2};-\kappa^{-2}\right)}{(N+2)\,\kappa^2}\;.
  \end{split}
\end{equation}
By construction, only the soft enhanced terms differ from the DGLAP
result. The Altarelli-Parisi splitting functions would give $\Gamma(N)=-2H_N$,
with $H_N$ the $N$th harmonic number, and ${\rm K}(N)=1/N$~\cite{Ellis:1991qj}.

\section{Momentum mapping and \texorpdfstring{\boldmath$q_T$}{qT} spectra in Drell-Yan type processes}
\label{sec:dy_qt}
This section presents a brief phenomenological analysis of the different recoil strategies 
described in Apps.~\ref{sec:css_kin_if} and~\ref{sec:css_kin_if_res}. We investigate the impact 
on the transverse momentum ($q_T$) spectrum of the Drell-Yan lepton pair at the LHC. 

It was pointed out~\cite{Nagy:2009vg} that the non-singlet initial-state parton evolution is generated 
in dipole-like parton showers by initial-state splitters with final-state spectator, except for 
the first branching, which stems from an initial-state splitter with initial-state spectator.
The kinematics mapping described in App.~\ref{sec:css_kin_if} then results in the Drell-Yan lepton pair
acquiring its entire transverse momentum in the first branching\footnote{
  Note that this argument is incomplete, as it is based on non-singlet parton evolution only.
  The leading real-emission configurations at center-of-mass energies much larger than 
  the di-lepton mass are not the $q\bar{q}\to l\bar{l}g$ channels, but $qg\to l\bar{l}q$.
  These processes are enhanced by a large gluon PDF. Similarly, the leading double
  real-emission configurations are $gg\to l\bar{l}q\bar{q}$. Both types of configurations 
  contain an initial-initial dipole, which -- when radiating additional partons -- generates
  recoil on the full final state, including the di-lepton pair.}. Comparing to analytical resummation
in impact parameter space~\cite{Collins:1981uk,*Collins:1984kg,*Collins:1985ue,*Sterman:1986aj,
  *Collins:1988ig,*Collins:1989gx}, one is led to conclude that only leading logarithms can be
resummed in this scheme~\cite{Parisi:1979se}. However, comparing to resummation in transverse
momentum space~\cite{Ellis:1997ii}, the modified leading logarithmic structure characteristic for 
parton showers emerges from Eq.~\eqref{eq:shower_functional}. A kinematics mapping more appropriate 
for comparison with analytical approaches is given by App.~\ref{sec:css_kin_if_res}. 
Here we restrict ourselves to analyzing its impact on the $q_T$ spectrum at the LHC. 
Similar analyses have been performed elsewhere for a variety of other observables using standard 
dipole-like parton showers based on Catani-Seymour subtraction~\cite{Hoeche:2009xc,Carli:2009cg}.

\begin{figure}[p]
  \centering
  \includegraphics[width=7.5cm]{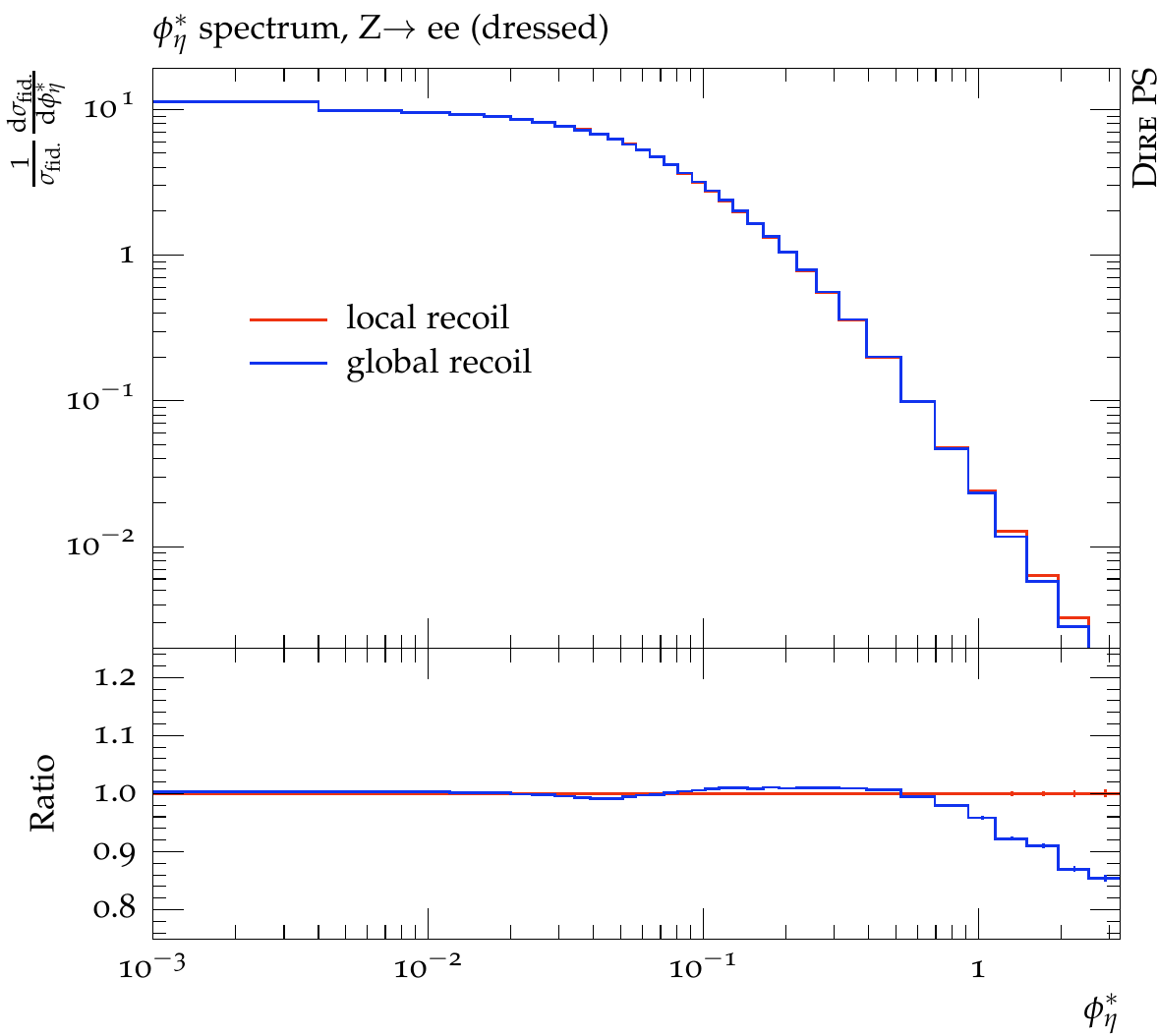}
  \includegraphics[width=7.5cm]{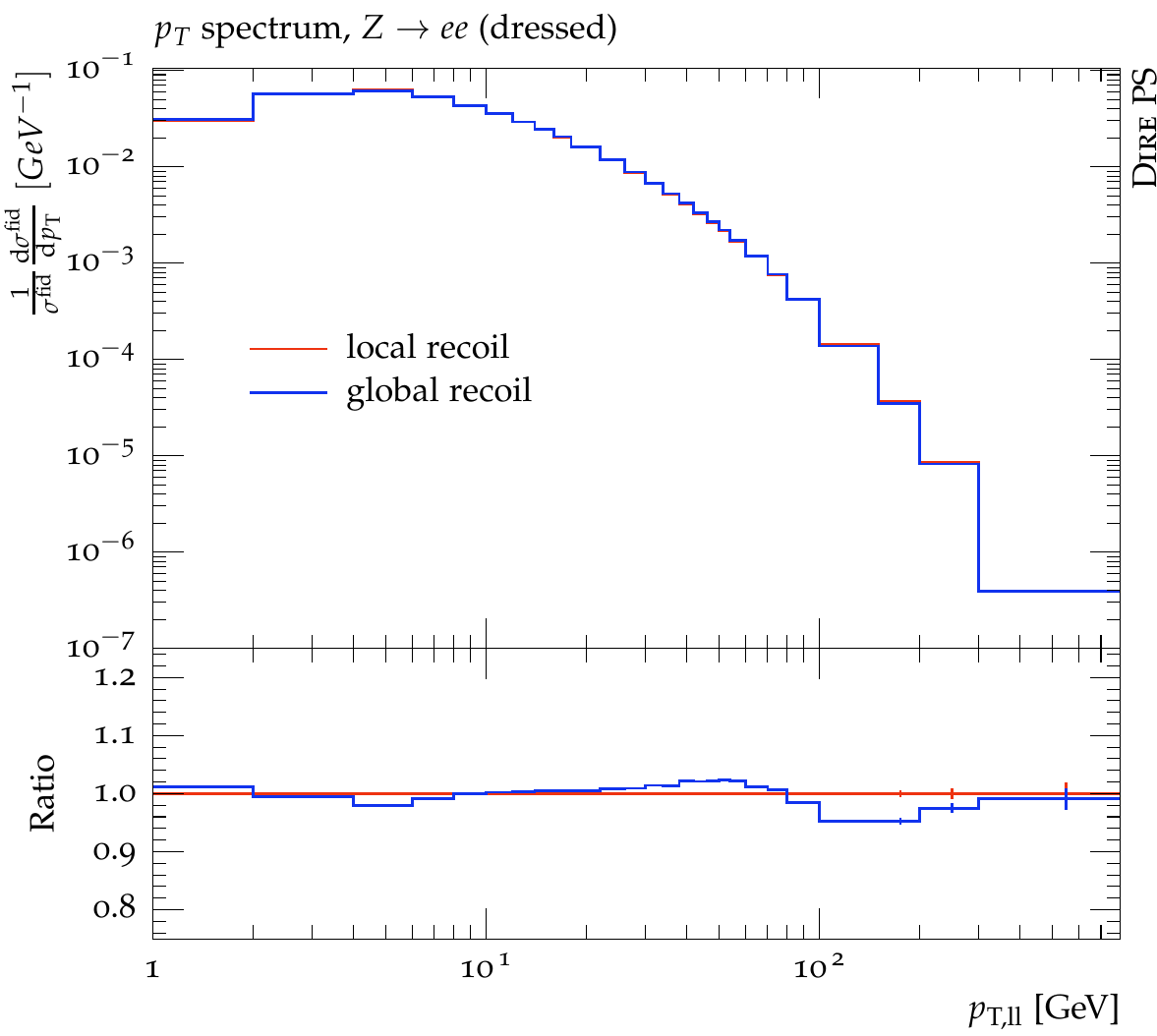}\\
  \begin{minipage}{7.5cm}
    \includegraphics[width=\textwidth]{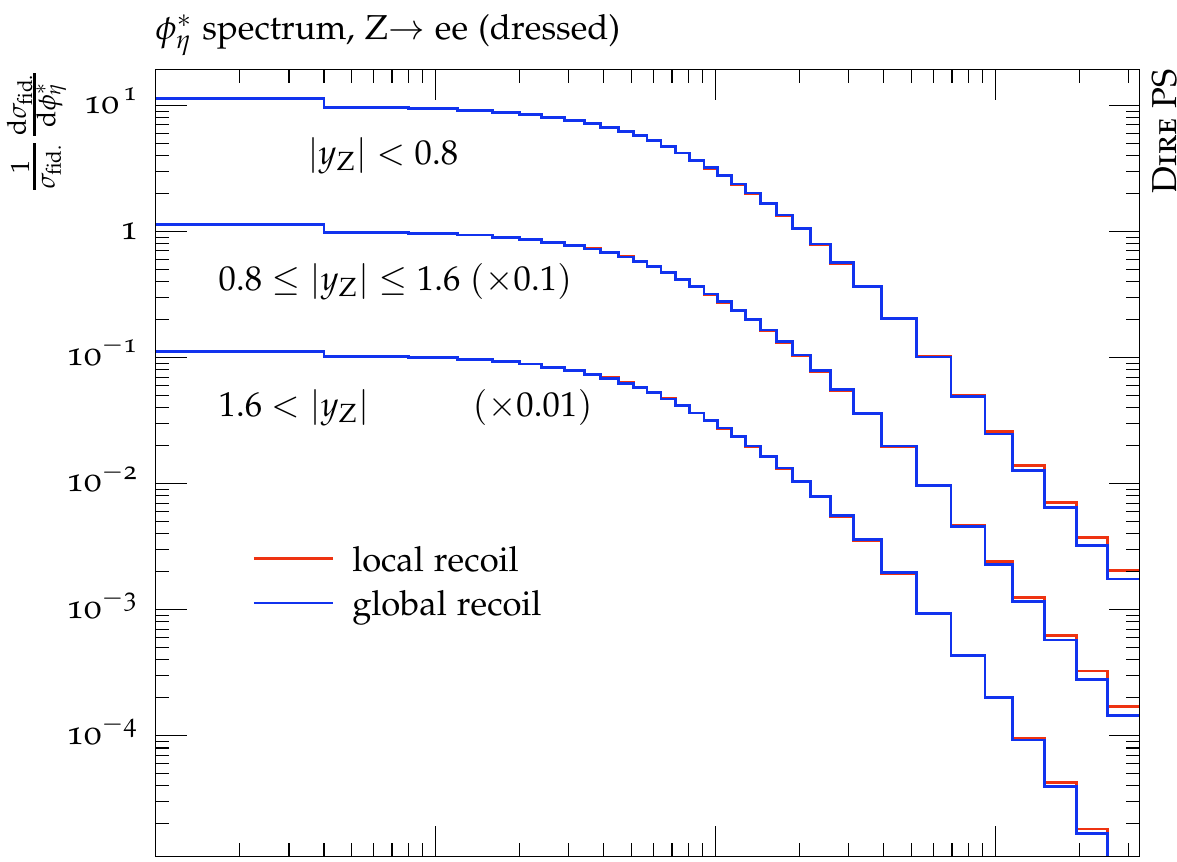}\\[-1mm]
    \includegraphics[width=\textwidth]{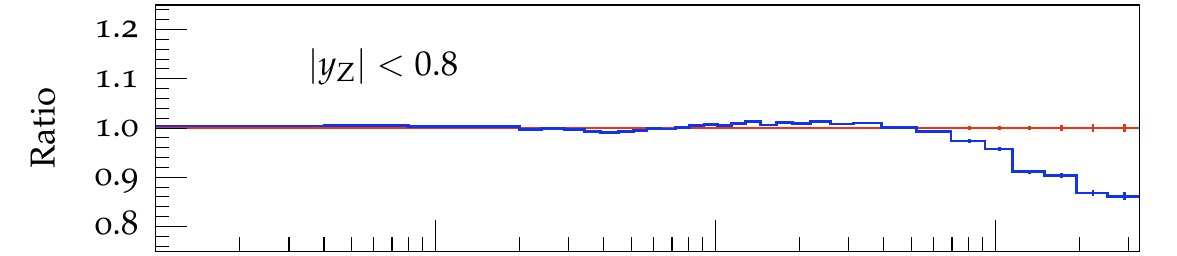}\\[-1mm]
    \includegraphics[width=\textwidth]{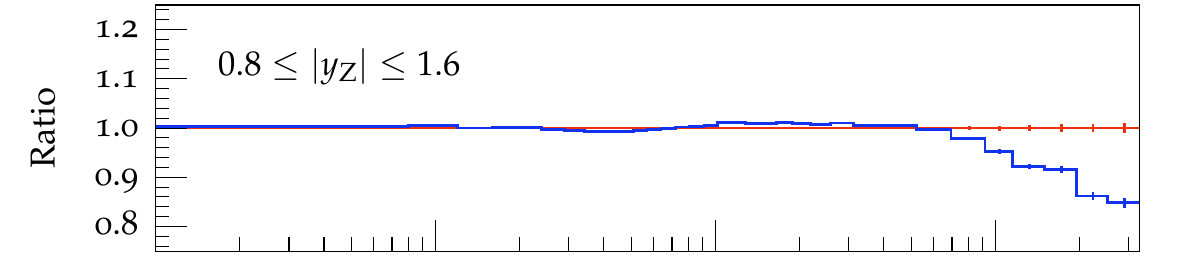}\\[-1mm]
    \includegraphics[width=\textwidth]{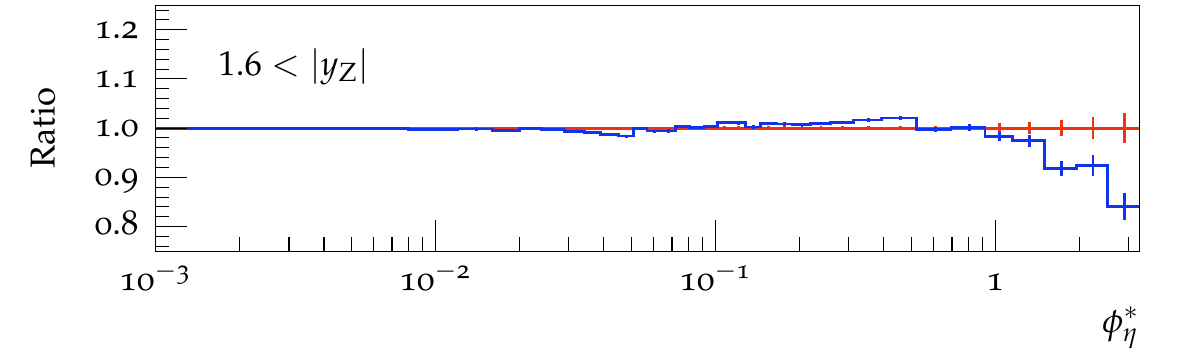}
  \end{minipage}
  \begin{minipage}{7.5cm}
    \includegraphics[width=\textwidth]{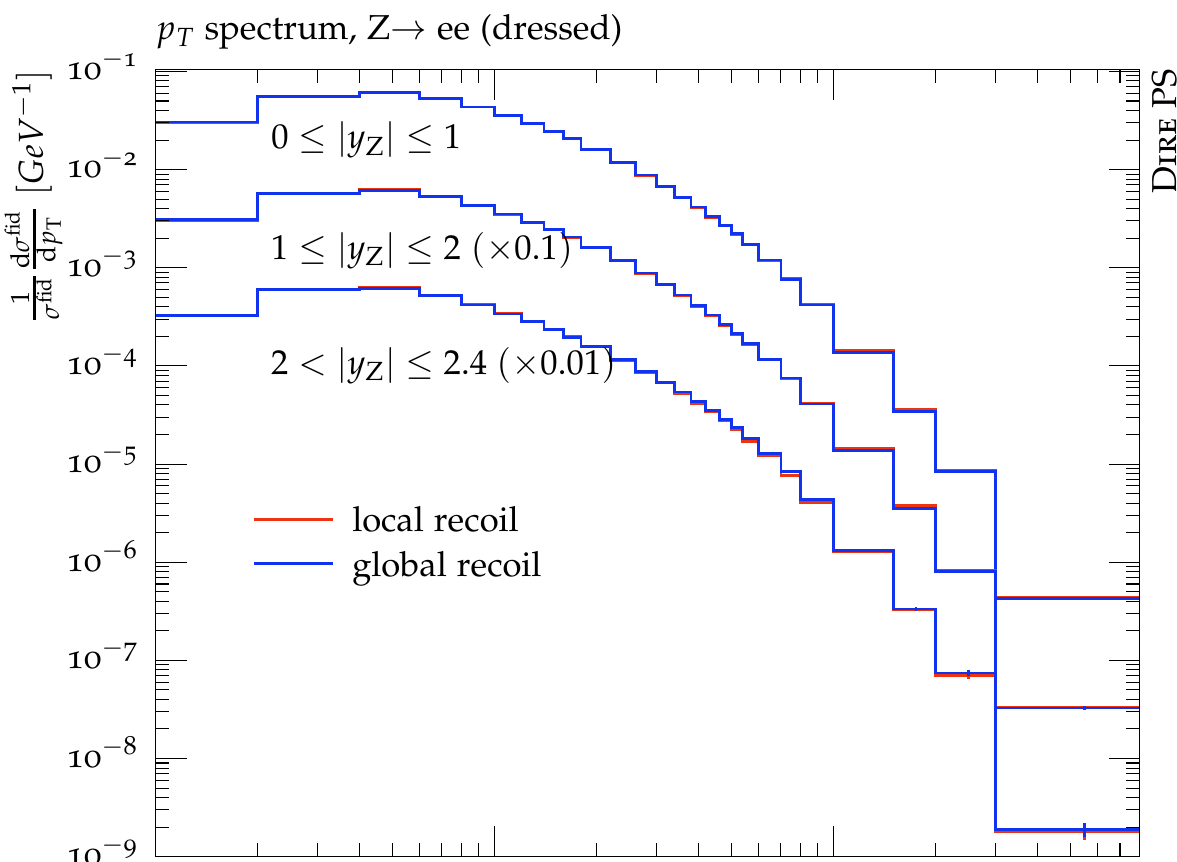}\\[-1mm]
    \includegraphics[width=\textwidth]{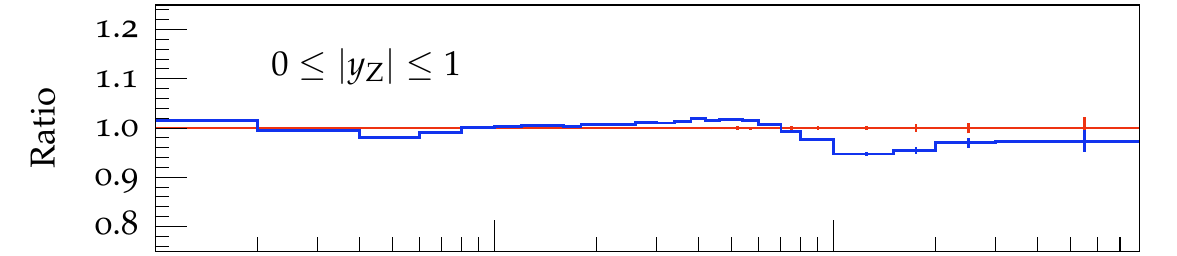}\\[-1mm]
    \includegraphics[width=\textwidth]{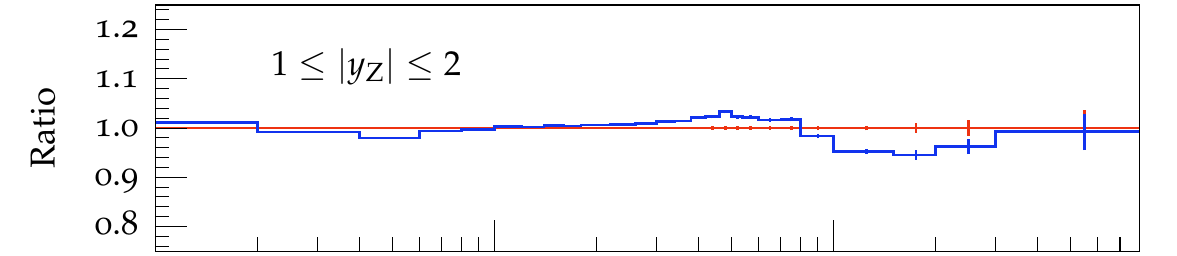}\\[-1mm]
    \includegraphics[width=\textwidth]{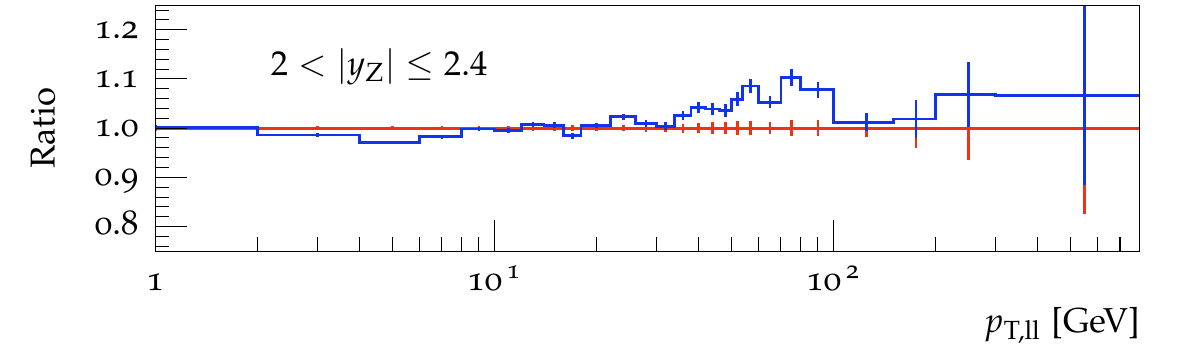}
  \end{minipage}
  \caption{
    The impact of the kinematics mapping on the $q_T$ spectrum of the Drell-Yan lepton pair at LHC energies.
    See Fig.~\ref{fig:lhc_ptspectra} for details on the analysis. We compare the local recoil scheme 
    from App.~\ref{sec:css_kin_if} with the global recoil scheme from App.~\ref{sec:css_kin_if_res}.
    \label{fig:qt_kinski}}
\end{figure}

Figure~\ref{fig:qt_kinski} shows \Dire predictions from \Sherpa for the two different kinematics schemes
described in Apps.~\ref{sec:css_kin_if} and~\ref{sec:css_kin_if_res}. To highlight the differences
in the resummation, we present pure parton shower results. Correspondingly, experimental data are omitted.
The global recoil scheme from \ref{sec:css_kin_if_res} shows a small difference at low $q_T$.
Similar conclusions were reached in~\cite{Hoeche:2009xc,Carli:2009cg}.

\clearpage
\bibliographystyle{amsunsrt_modp}
\bibliography{journal}
\end{document}